# Approach to Hyperuniformity in a Metallic Glass-Forming Material Exhibiting a Fragile to Strong Glass Transition

*Hao Zhang[1†], Xinyi Wang[1], Jiarui Zhang[1], Hai-Bin Yu[2], Jack F. Douglas[3†]*

[1] Department of Chemical and Materials Engineering, University of Alberta, Edmonton, Alberta, Canada, T6G 1H9

[2] Wuhan National High Magnetic Field Center, Huazhong University of Science and Technology, Wuhan, Hubei, China, 430074

[3] Material Measurement Laboratory, Material Science and Engineering Division, National Institute of Standards and Technology, Maryland, USA, 20899

*Corresponding authors: hao.zhang@ualberta.ca; jack.douglas@nist.gov

**Abstract**

We investigate a metallic glass-forming (GF) material ($Al_{90}Sm_{10}$) exhibiting a fragile-strong (FS) glass-formation by molecular dynamics simulation to better understand this highly distinctive pattern of glass-formation in which many of the usual phenomenological relations describing relaxation times and diffusion of ordinary GF liquids no longer apply, and where instead genuine thermodynamic features are observed in response functions and little thermodynamic signature is exhibited at the glass transition temperature, $T_g$. Given the many unexpected similarities between the thermodynamics and dynamics of this metallic GF material with water, we first focus on the anomalous static scattering in this liquid, following recent studies on water, silicon and other FS GF liquids. We quantify the "hyperuniformity index" $H$ of our liquid, which provides a quantitative measure of molecular "jamming". To gain insight into the $T$-dependence and magnitude of $H$, we also estimated another more familiar measure of particle localization, the Debye-Waller parameter $\langle u^2 \rangle$ describing the mean-square particle displacement on a timescale on the order of the fast relaxation time, and we also calculate $H$ and $\langle u^2 \rangle$ for heated crystalline Cu. This comparative analysis between $H$ and $\langle u^2 \rangle$ for crystalline and metallic glass materials allows us to understand the critical value of $H$ on the order of $10^{-3}$ as being analogous to the Lindemann criterion for both the melting of crystals and the "softening" of glasses. We further interpret the emergence of FS GF and liquid-liquid phase separation in this class of liquids to arise from a cooperative self-assembly process in the GF liquid.

**I. Introduction**

In recent papers [1-2] investigating the nature of the Johari- Goldstein (JG) β-relaxation process and other aspects of the "fast dynamics" of a model Al-Sm metallic glass-forming (GF) liquid, we found that this material exhibited an unexpected fragile-to-strong (FS) glass transition having many similarities to glass-formation in water, silica and other network forming liquids. Even though this complicated our analysis of the temperature ($T$) dependence of diffusion, structural relaxation, the nature of dynamic heterogeneity, and collective motion in these materials, we found that many aspects of these materials could be understood based on the same framework utilized in quantifying these properties in "ordinary" glass-forming (OGF) liquids. In particular, we found that the dynamic heterogeneity in this metallic GF liquid is *identical* to that found previously in Cu-Zr metallic glass materials [3], which exhibit "ordinary" glass-formation. We also found that the string model of relaxation in GF liquids [19, 20] quantitatively describes mass diffusion in this FS GF system over a large range of temperatures so that this FS GF liquid fits into a pattern of dynamics observed previously, although the $T$ dependence of the dynamic heterogeneity is somewhat different, explaining the difference in relaxation time and diffusion in these different classes of glass-formers (In our discussion below, we will summarize some of the defining features of FS glass-formation and some of our most important findings for the Al-Sm material.).

The present paper is aimed at better understanding the origin of the FS glass-formation. Another goal is also to develop a unified conceptual framework for understanding the dynamics of GF liquids that encompasses both fragile-strong and ordinary glass-formation. One characteristic feature of water, silica and other GF liquids undergoing FS glass-formation is that these liquids seem to have exceptionally low isothermal compressibility values at low $T$, and this prompted us to study the "hyperuniformity index" $H$, [4] a dimensionless measure of molecular

4"jamming" defined in terms of the isothermal compressibility. To better comprehend general trends and critical values of this dimensionless jamming measure, we also calculated the mean square atomic displacement $\langle u^2 \rangle$ on the time scale of the fast β-relaxation, which is typically on the order of a ps in molecular fluids. Of course, the Debye-Waller parameter $\langle u^2 \rangle$ can be measured by a variety of methods and can readily be estimated by simulation and this quantity is certainly a more familiar quantity than $H$. Through a comparative analysis of $\langle u^2 \rangle$ and $H$ we can ascribe a define physical meaning to the "critical hyperuniformity index", $H_c \sim O(10^{-3})$, defining the emergence of materials in an "effectively hyperuniform" state. We find this relation can be understood as a kind of generalized Lindeman criterion. We also show through an explicit computation that $H$ is non-zero in a model crystalline material (crystalline Cu) at finite temperatures, as in the case of $\langle u^2 \rangle$, and that $H$ increases in our model crystalline material, as $H$ also does in GF materials at low $T$. Together, these calculations clarify the meaning of $H$ and characteristic values of this "jamming" index in both fluids and solids in their equilibrium state. While our analysis of $H$ shows that this quantity leads to relatively low $H$ values seen in previous simulations of water, silicon and silica (See discussion below), this observation still does not explain why the values of $H$ are apparently exceptionally low in fluids exhibiting FS GF. To address this question, we explore a tentative hypothesis that dynamic polymerization is universal to all GF liquids, but that the highly cooperative nature of branched equilibrium polymerization occurring in liquids exhibiting FS GF in comparison to linear chain polymerization occurring in "ordinary" GF liquids accounts for many of the distinct properties of these classes of GF liquids, including the propensity towards liquid-liquid phase separation even in single component FS GF liquids.

At the outset, we should acknowledge that our idea of studying the emergence of hyperuniformity in our metallic glass material was greatly influenced by recent computational studies of water, [5-6] silicon, [7-8] and silica. [9] Evidently, the propensity to approach effective hyperuniformity at low $T$ is a common if not universal attribute of solidification of FS GF materials. We then wondered whether our Al-Sm metallic glass had this property.

Our work was also motivated by recent work indicating that an approach to hyperuniformity was also characteristic of polymer grafted nanoparticles having moderate cross-linking density where there are large fluctuations of the polymer segment density in the grafted polymer layer. [10-11] In particular, we hypothesized that the large configurational polarizability of polymer grafted nanoparticles[12] might be physically analogous to the relatively large polarizability of the many-electron Lanthanide Sm atoms in our metallic glass material (We are not aware of any precise estimate of the polarizability of Sm, but the polarizability of elements normally increases roughly linearly with atomic volume [13] so on this basis we expect the polarizability of Sm to be relatively large in comparison with more common metallic elements.). Finally, we mention the interesting work of Sciortino and coworkers [14] devoted to coarse-grained models of water, silicon and silica in which the intermolecular potentials are modelled in terms of patchy colloid models that show an inherent tendency to form tetrahedral networks and to approach hyperuniform state at low $T$. Other simulation studies on this type of patchy colloid model (the patches being both particle-like or described by grafted polymer chains) have also shown an inherent tendency of the coarse-grained fluids having sticky "spots" exhibit multiple critical points and the phenomenon of liquid-liquid phase separation [15-18], common properties of liquids undergoing FS GF, as we discuss below. These simulation studies collectively suggest the formation of a dynamic network structure might be an essential feature of glass-forming liquids undergoing FS GF, and below argue that it is just this feature of liquids undergoing FS GF that gives rise to an approach to their effective hyperuniformity.

**II. Model and Simulation Methods**

Our molecular dynamics (MD) simulations of the $Al_{90}Sm_{10}$ metallic GF alloy are based on a many-body potential developed by Mendelev et al. [19] This potential is of the Finnis-Sinclair type [20] and is semiempirical in nature because the parameters in this model were determined to optimize the consistency of the model calculations for the cohesive energy density, elastic modulus, vacancy formation energy, melting point of pure aluminum. In addition to the capacity of reproducing many of the properties of pure Al materials, it has the added advantages of providing excellent formation energies for a series of Al-rich crystal phases and provides excellent reproduction of the measured structure factor of the material investigated in the present paper at high $T$ and good agreement with ab initio MD simulations revealing dominating short-range-order corresponding to an Sm-centered motif lead us to expect this model to be suitable for simulating this alloy. [19, 21] Another important attribute of this metallic glass model is that it is highly resistant to crystallization, which is necessary for simulations extending to very low $T$ where relaxation times become very long.

The simulated material was composed of 28785 Al atoms and 3215 Sm atoms and the $T$ was initially held at 2000 K for 2.5 ns in order to reach an apparent equilibrium. The liquid then was cooled continuously to 200 K with a cooling rate of 0.1 K/ns. Despite using a very slow cooling rate in this study, the system will not be able to reach full equilibrium at low temperatures because the cooling time is shorter than the relaxation time. Periodic boundary conditions were applied in all directions and an isobaric-isothermal ensemble (NPT) was employed where $P = 0$. The simulation box size was controlled using the Parrinello-Rahman method [22] and $T$ was maintained by a Nose-Hoover thermostat. [23-24] The MD simulations utilize Large-scale Atomic/Molecular

Massively Parallel Simulator (LAMMPS) [25], developed at the Sandia National Laboratories. We also isothermally heated the material for an extended period of time to enhance the equilibration of the material and to probe kinetic processes that cannot be observed under continuous heating conditions. Isothermal heating simulations were performed for a range of temperatures: $T$ = 900 K, 850 K, 800 K, 750 K, 700 K, 650 K, 600 K, 550 K, 500 K and 450 K. The simulations were performed for at least 10 ns and up to 0.7 μs, where the simulation time is chosen to be longer than the structural relaxation time at a $T$ above $T_c$ (defined below) to ensure the system reaches equilibrium.

As a comparison to our metallic GF liquid simulations, we also consider a model crystalline Cu material in which the atomic interaction between Cu atoms is described by a widely used embedded atom model (EAM) potential developed by Mishin et al. [26] A perfect face-centred cubic (FCC) Cu crystal of 13,500 atoms with periodic boundary conditions in all directions was heated from 200 K to 1800 K with a heat rate of $10^{11}$ K/s, until the crystal was totally melted. Isothermal heating simulations were also performed at $T$ = 1500 K, 1350 K, 1200 K, 900 K, 600 K, and 300 K with a canonical ensemble. At each $T$, the simulation was conducted for 1 ns.

### III.  Results and Discussion

In our previous studies [1-2], which were mainly focussed on the physical nature of the Johari-Goldstein $\beta$-relaxation and fast relaxation processes in a model Al-Sm metallic GF fluid, we observed a direct correspondence between the JG $\beta$-relaxation time $\tau_{JG}$ and the lifetime of the mobile particle clusters $\tau_M$. We also established a direct relation between $\tau_{JG}$ and the rate of molecular diffusion $D$ in this material in previous work [1-2], which is practically important because the JG $\beta$-relaxation process becomes the prevalent mode of relaxation in materials in their glass state. These findings complement earlier observations of a direct relationship between the immobile particle cluster lifetime $\tau_I$ and the average structural relaxation time $\tau_\alpha$, obtained from the decay of the intermediate scattering function. [3, 27] The general picture indicated by these previous works [3, 27-28] is that dynamic clusters of mobile particles dominate the rate of diffusion, while clusters of immobile particles dominate the rate of structural relaxation. The disparity between the lifetimes of the mobile and immobile particle clusters then accounts for the "decoupling" phenomenon between the rate of mass diffusion and structural relaxation in GF materials, a phenomenon that tends to be amplified at lower $T$. [1] This aspect of GF liquids appears to be general for *all* GF liquids.

**A. Equilibrium and Structural Properties**

The equilibrium thermodynamic, structural, and rheological characteristics of our model metallic GF liquid have been rather thoroughly investigated previously to "validate" the interatomic potential, based on experimental consistency criteria. For example, previous work has shown that the "structure" of this metallic GF liquid, based on the pair correlation function, is well reproduced at $T$ = 1273 K. We expected this potential to provide a realistic description of the material in its glass state since this potential reproduces the short-range order of this material predicted by ab initio molecular dynamics simulations (AIMD). [29] Another aspect of this model that has been rather exhaustively investigated is the tendency of the atomic species to form locally icosahedral-packed structures in the liquid and for these domains to form extended polymeric structures upon approaching the glass transition. [21] This tendency mesoscale local structure formation in the form of "strings" (i.e., structures having a polymeric geometrical form) of icosahedral atomic clusters is also highly prevalent in Cu-Zr and other metallic glass forming materials exhibiting OGF [30-31] so there is nothing particularly unique about this form of local ordering in our $Al_{90}Sm_{10}$ metallic GF alloy. We conclude from this extensive prior analysis that there does not appear to be anything "special" about the $Al_{90}Sm_{10}$ metallic GF liquid from a structural standpoint that might obviously explain the non-standard pattern of the dynamics in this class of GF materials. We then apparently need to look elsewhere for the origin of FS glass-formation in this material.

One of the characteristics of crystalline materials that differentiate them from most GF materials is that they tend to have relatively small isothermal compressibility. Indeed, the isothermal compressibility becomes small in crystalline materials as $T$ approaches zero where the treatment of crystalline materials in terms of lattices of particles interacting with harmonic interactions becomes a good approximation. At finite $T$, however, crystals and other real materials, undergo thermal expansion due to emergent anharmonic interactions between the molecules and the isothermal compressibility can become appreciable (We quantify this phenomenon below for a model crystalline Cu material to get a "feel" for the relative magnitudes involved.). The larger amplitude thermal motions of the particles in the expanded lattice also give rise to a $T$-dependent shear and bulk moduli, but the isothermal compressibility, the reciprocal of the bulk modulus, is characteristically rather small in most crystalline materials.

Recent works by Torquato and Stillinger [4, 32-33] have indicated that some strongly interacting fluids, especially those with soft intermolecular interactions, as found in dusty plasmas [34-35] and nanoparticles with grafted polymer layers [10, 36] having a similar hard core and soft shell repulsive interparticle interaction, exhibit exceptionally low isothermal compressibility values in comparison to "normal" liquids. This is apparently the case also for water and silica, which are notably liquids that exhibit FS glass-formation (We discuss these fluids and aspects of this type of glass-formation below.). In physical terms, the existence of relatively low isothermal compressibility means that the molecules or other particle species in the fluid are strongly

"hemmed in", or "jammed" in the colloquial sense of this term, based on this long wavelength thermodynamic criterion. (To avoid potential confusion, we note that we are not referring here to the narrower technical definition of "jamming" discussed by Torquato [33] and others.). We may also define a measure of *local jamming* from the height of the first peak of the static structure factor $S(q)$ of the liquid (Fourier transform of the pair correlation function) where a higher peak height indicates stronger interparticle correlations at the length scale of the interparticle distance that derive either from the action of stronger repulsive interatomic excluded volume interactions or attractive cohesive intermolecular interactions. Notably, the condition at which the height $S_p$ of the primary peak in the structure factor $S(q)$ reaches a "critical" value has often been taken as a phenomenological criterion for the onset of fluid "freezing". In particular, the Hansen-Verlet freezing criterion [37-40] corresponds to $S_p$ being in a range between 2.85 to 3 (This onset condition also appears to roughly locate the onset of non-Arrhenius relaxation dynamics in model GF polymer liquids.[41]). These jamming measures at macroscopic and molecular scales can be brought together by defining a dimensionless ratio, the "hyperuniformity index" [4], $H \equiv \lim_{q \to 0^+} S(q)/S_p$, which provides a well-defined, and often an experimentally accessible measure of the extent of "jamming" in condensed materials in the sense described above.

Material systems having a value of $H$ less than a value on the order of magnitude $10^{-3}$ have been previously defined to be "effectively hyperuniform" [5], and such materials have been of great recent interest because of a wide range of predicted material properties of this class of materials. [4] Comparably low extrapolated values of $H$ have been observed in simulations of coarse-grained simulations of polymer melts [41] at $T$ appreciably below the estimated $T_g$ of these materials where it was suggested that $H$ approaching on the order of $10^{-3}$ near the extrapolated Vogel-Fulcher-Tammann temperature, the $T$ at which the structural relaxation time correspondingly extrapolates to infinity in the VFT equation. However, further work is required to better understand the physical significance of the suggested critical value of $H \approx 10^{-3}$ defining the emergence of "effective hyperuniformity". We discuss this fundamental question below as part of our investigation of the origin of FS glass-formation in our Al-Sm GF liquid.

At this point, it is notable that both water [42-44] and silica [45] have been observed to exhibit an FS transition in their dynamics, as well as emergent hyperuniformity in their amorphous solid ("glass") states. Since we also observe an FS transition in our $Al_{90}Sm_{10}$ GF liquid, and because of the prevalence of this type of glass-formation in other metallic and non-metallic materials [46,47], it is then of evident interest to consider $H$ in our metallic GF liquid simulations. We anticipated this phenomenon might arise in this metallic GF material because of the relatively "soft" interatomic interactions of the relatively heavy and polarizable Sm atoms. "Soft" interactions are a common feature of many approximately hyperuniform real materials because such interactions allow greater particle penetration into the domains of surrounding particles, and thus should lead to stronger jamming under high particle density conditions. [10]

In Figure 1, we show the $T$ variation of our Sm-Al metallic glass material. The structure factors are calculated by the Fourier transform of the pair correlation function,

$$S(q) = 1 + 4\pi\rho \int \frac{sinqr}{qr} g(r) r^2 dr \qquad (1)$$

where $r$ is number density, $g(r)$ is the radial distribution function. [48-49] The primary peak of the structure factor $S(q)$ grows progressively upon cooling, reflecting the local jamming of molecules and we see that $\lim_{q \to 0^+} S(q)$ becomes progressively smaller as the $T$ is lowered. These trends together imply that $H$ is decreasing upon cooling, and we show our estimates of $H$ and $S_p$ as a function of $T$ in the inset of Fig. 1. $H$ indeed progressively decreases upon lowering $T$, approaching a nearly hyperuniform condition upon cooling (This trend is quantified below.). We discuss the $T$ dependence of $H$ in greater detail below in relation to findings made in earlier work for the characteristic temperatures of glass-formation and other properties closely related to $H$ to better understand the trend indicated in Fig. 1.

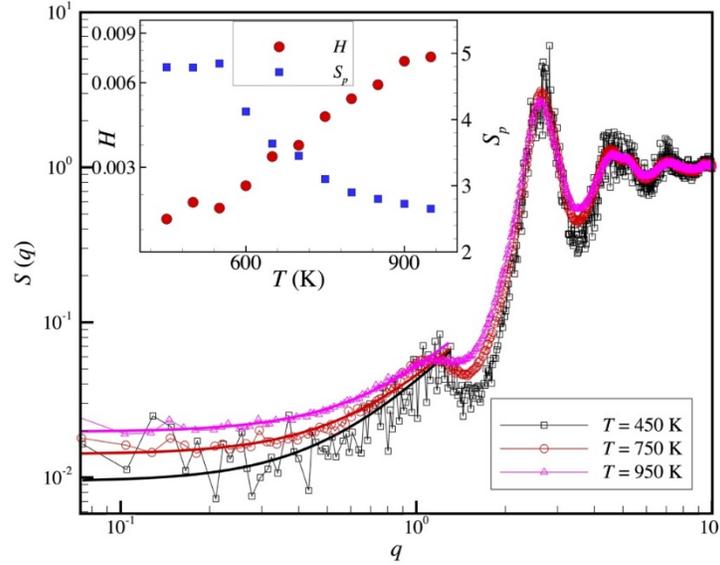

**Figure 1.** Static structure factor $S(q)$ of $Al_{90}Sm_{10}$ metallic glass forming liquid for different $T$. Inset shows the hyperuniformity parameter, the ratio of $S(q)$ extrapolated to its long wavelength ($q \to 0^+$) limit to the height $S_p$ of the principal maximum in $S(q)$ over the entire $T$ range investigated. The bold solid lines are a second-order polynomial fitting of $S(q)$ in the low $q$ range, following Huang et al. [50].

We also observe a low-$q$ upturn in $S(q)$ of our metallic glass, as in the case of water, [50] which suggests the development of some type of large-scale structure formation in this class of GF liquids. A similar low $q$ upturn in $S(q)$ is also apparent in measurements on dusty plasmas, as recently analyzed by Zhuravlyov et al. [51], and, indeed, an upturn of this kind is observed in many GF liquids. [52-54] In addition to this common, but not universally observed, upturn effect in $S(q)$, we also see a small "pre-peak" in a lower $q$ than the main peak of the structure factor at which $S_p$ is defined. This regime is often termed a mesoscale regime because it is intermediate between the size of the molecules and the macroscopic scales of the bulk material. A pre-peak also arises in Cu-Zr(Al) metallic glasses. [55] The interpretation of these "anomalous" scattering features is a complex and controversial problem, and we discuss the possible origin of these unexpected scattering features in the Supplementary Information section of the paper since this topic is peripheral to the main topic of our paper. The primary problem of the present paper is understanding the *T*-dependence of *H* and its possible relevance for understanding FS glass formation generally.

At this point, some explanation is required for how we estimate the hyperuniformity index *H*, given the low-*q* upturn in Fig. 1, which evidently complicates the estimation of *S(0)*. We approached this problem in the same way as in previous X-ray scattering measurements on *S(q)* in water at low temperatures where a similar upturn conspicuously arises. [50] Huang et al. [50] fitted their *S(q)* in the *q* range well below the minimum in water to second-order in a polynomial in *q* and then extrapolated their data to the thermodynamic limit, $q = 0$. Based on this procedure, Huang et al. found that the result of this procedure was quantitatively consistent with isothermal compressibility estimates independently obtained from earlier sound velocity measurements, thus validating their extrapolation procedure. We estimate *S(0)* for our Al-Sm metallic glass following this same procedure, and these results, in conjunction with those for the principal peak height of *S(q)*, allow us to estimate the values of *H* shown in Fig. 1. While estimates of *H* are somewhat uncertain because of the limited computational size of our simulations and the necessity of this extrapolation procedure, the general trend seems to be clear. We point out that we found numerous commonalities between the thermodynamics and dynamics of cooled water and our Al-Sm metallic glass in our previous studies, [1-2] so we find the similarity between the static scattering between these fluids at low temperatures to be quite natural. We next try to better understand the *T*-dependence through a consideration of other more familiar properties related to the fluid isothermal compressibility. We are concerned with the persistent question of why a critical value of *H* should exist upon approaching an amorphous solid state. Since crystalline materials provide a well-known class of solid materials, we also consider the magnitude of *H* in crystalline Cu over a wide range of *T* to compare with the *H* estimates in a metallic glass approaching "solidification". We find that *H* is indeed finite in our model crystalline material and this quantity increase monotonically with *T* in a way similar to our model GF liquid, although the magnitude of *H* tends to be significantly smaller than in the metallic GF liquid.

## *B. H* and Alternative Jamming Measures

As discussed above, the hyperuniformity index *H* is a quantitative measure of particle jamming at large length scales relative to atomic scales so it is natural to look for other properties of liquids which encode similar information, and which are more familiar and experimentally accessible, that might help us better understand the *T* variation of *H* in Fig. 1. Recently, it has become appreciated that the mean square particle displacement on a timescale comparable to the fast relaxation time (typically a timescale on the order of a ps in liquids), the Debye Waller parameter $\langle u^2 \rangle$, bears a strong correlative relationship to the fluid bulk modulus *B*, the reciprocal of the isothermal compressibility, $B \sim \langle u^2 \rangle^{3/2}$.[56-59] Specifically, $\langle u^2 \rangle$ is defined as,

$$\langle u^2 \rangle = \langle \frac{1}{N} \sum_{n=1}^{N} \{(x_1 - x_0)^2 + (y_1 - y_0)^2 + (z_1 - z_0)^2\} \rangle, \tag{2}$$

where $(x_0, y_0, z_0)$ and $(x_1, y_1, z_1)$ are particle's initial and final positions after time $t = 1$ ps, respectively,[3, 60-61] which is notably an equilibrium fluid property.

This commonly measured "fast dynamics" property has also been found to be a good measure of material stiffness at the scale of the size of the particles.[56-59] $\langle u^2 \rangle^{3/2}$ defines the average volume explored by the center of the particles in their cage created by the presence of surrounding particles, and thus reflects a combination of structural constraints arising from repulsive excluded volume interactions between the particles, as embodied by static free volume ideas of fluids, and a contribution arising from the kinetic energy of the particles that act to opposes these constraints. The kinetic energy contribution ultimately gives rise to the "fluidity" of liquids under thermodynamic conditions and should not be neglected when considering the volume accessible to a particle in a fluid. We then see that Debye-Waller parameter $\langle u^2 \rangle$ defines a dynamical variety of "dynamical free volume". The compressibility likewise reflects a competition between this excluded volume and inertial effects arising from the kinetic energy of the particles in materials at equilibrium. Given this qualitative interpretation of $\langle u^2 \rangle$, it is natural to estimate this quantity as a function of $T$ to see if it provides any insight into the temperature variation of $H$.

At the outset, we note that melting in crystalline materials and the "softening" in glass materials have often been correlated with critical values of $\langle u^2 \rangle$, i.e., the empirical Lindemann criterion, [56, 62-63] so we might hope that critical values of $H$ might have a similar interpretation. The phenomenological Hansen-Verlet condition for freezing in terms of a critical value of $S_p$ also points in this direction. We next summarize our findings for $\langle u^2 \rangle$ in the Al-Sm system, and then move on to consider the $T$-dependence of $H$.

In the inset of Figure 2, we plot $\langle u^2 \rangle$ over the full $T$ range that we have investigated. We notice that as usual, there is a low $T$ regime and that $\langle u^2 \rangle$ extrapolates to zero at a finite $T$, defining a characteristic temperature, $T_o = 190$ K, indicating the "termination" of the glass-formation process. It should be appreciated that this $T$ involves a *long extrapolation* and should not be literally interpreted as the $T$ at which $\langle u^2 \rangle$ vanishes, given that our data is limited to a high $T$ regime. Within the Localization Model of glass-formation in which the structural relaxation time is related to $\langle u^2 \rangle$,[61] the characteristic temperature $T_{o,u} = 190$ K corresponds to the same $T$ at which the structural relaxation time $\tau_\alpha$ extrapolates to ∞. A formal divergence of this kind also arises in the well-known VFT equation [64-66], which normally describes the structural relaxation time and diffusion data for OGF liquids over a large $T$ range *above* $T_g$, but below the crossover temperature $T_c$, defined below. [67-68] Again, the divergence of the relaxation time cannot be taken literally because of the long extrapolation involved in estimating $T_o$.

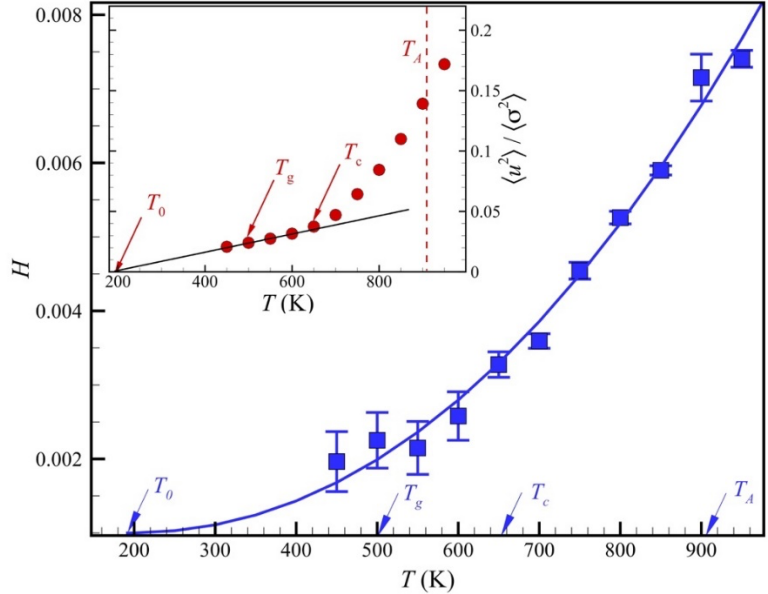

**Figure 2.** Comparison of the hyperuniformity index $H$ and the Debye-Waller parameter $\langle u^2 \rangle$ over a wide range of temperatures for our Al-Sm metallic glass where the characteristic temperatures of glass-formation are indicated. The uncertainty estimate is based on a 95% confidence interval. The $T$ dependence of $\langle u^2 \rangle$ (see figure inset) is rather typical of GF materials generally where a low $T$ near linear $T$ dependence is observed where $\langle u^2 \rangle$ extrapolates to zero near a characteristic temperature $T_o \approx 190$ K at which transport properties like the shear viscosity of structural relaxation time extrapolate to infinity, while at higher $T$ becomes strongly non-linear in its $T$ dependence, a change of behavior often attributed to the "anharmonicity" of the interparticle interactions. At low temperatures, we see that $H$ exhibits a non-linear $T$ dependence, $H \approx H_o + (3.5 \times 10^{-4}) [(T - T_o)/T_o]^{2.13}$ where $H_o = 1.0 \times 10^{-3}$ (shown as solid line in comparison to our $H$ estimates). $H_o$ just happens to correspond to the effective hyperuniformity condition discussed above. The fitted exponent near 2.13 is larger than the value of 1.5 expected from the expected scaling between the isothermal compressibility, $\kappa_T \sim \langle u^2 \rangle^{3/2}$ under conditions when $S_p$ is nearly independent of $T$, i.e., low $T$ where $\langle u^2 \rangle$ varies linearly with $T$. Characteristic temperatures for glass-formation are indicated for reference where their definition is described in the text and further discussed in our previous studies.[1]

It should be appreciated that the VFT equation *does not* describe relaxation time and diffusion data in the Al-Sm metallic GF system (see Sect. of the SI of Ref. 1) over a large $T$ range, which is the normal situation for fluids exhibiting FS glass-formation. Moreover, no simple functional form for the $T$ dependence of relaxation and diffusion in FS GF systems is currently known and these fluids provide a good test of any theory of GF liquids purporting to be general. We next describe the estimation of the other characteristic temperatures of glass-formation using the same methodology as for Zr-Cu metallic GF liquids. [61]

## C. Estimation of Characteristic Temperatures of Glass-formation from $\langle u^2 \rangle$ Data

A basic problem that arises in defining the glass transition temperature $T_g$ in materials exhibiting a fragile-to-strong (FS) glass-formation is that one cannot rely on the standard phenomenology of GF liquids such as the VFT relation [64-66], the position of the peak position in the specific heat as the temperature is varied or the condition at which the α-relaxation time $\tau_\alpha$ equals 100 s at this temperature. As an example, we consider the situation of water, the most extensively studied fluid exhibiting an FS transition. [42-43] Estimates of $T_g$ for water based on the "100 s rule" or its shear viscosity equivalent, in conjunction with the VFT relation, have indicated a $T_g$ estimate for water near (162 ± 1) K,[69-70] while other estimates, based on specific heat $C_p$ measurements of amorphous ice, have indicated a much lower value of $T_g$ near 136 K. [71] While the lower $T_g$ estimate seems to have a greater "acceptance" in the scientific literature, there is currently no general consensus on the value of $T_g$ for water, and one routinely finds both of these $T_g$ estimates reported as the glass transition temperature of the water. The study of glass-formation in water is complicated by the propensity of bulk water to crystallize at very low temperatures so this transition is often studied in confined water and by molecular dynamics simulation.

We address the problem of estimating the characteristic temperatures of glass-formation in our Al-Sm GF liquid based on a highly simplified and apparently robust method that allows estimation of the characteristic temperatures of glass-formation from the estimation of a series of critical conditions defined in terms of critical values of $\langle u^2 \rangle$. In effect, we define the characteristic temperatures of glass-formation in terms of generalized Lindemann-type criteria. Beforehand, we note that the resulting characteristic temperature estimates of GF coincide to a remarkable degree of approximation to estimates made based on much more elaborate calculations of the $T$ dependence of the intermediate scattering function and diffusivities of the atomic species in our simulated Al-Sm fluid. This discussion also provides us with an opportunity for reviewing some of the singular characteristics of FS type glass formation that distinguish this type of glass formation from OGF liquids. Another purpose of this discussion is to provide a useful metrology for the stages of glass formation, for this very different type of glass-formation even standard phenomenological equations of OGF liquids, such as the VFT equation, are no longer generally applicable. The success of this methodology of estimating the characteristic temperatures of glass-formation based on $\langle u^2 \rangle$ should be described in terms of a common theoretical framework, despite the many superficial differences between these general classes of GF liquids.

First, we estimate $T_g$ by a Lindemann estimate appropriate for a *fragile* glass-former, $\langle u^2 \rangle^{1/2} / \sigma \approx 0.15$ following the arguments of Dudowicz et al.[67] For the present system this criterion leads to the estimate, $T_{g,u}$ = 500 K. This $T_g$ estimate is in "reasonable" agreement with the estimate of the $T$ = 520 K at which structural relaxation and diffusion in our metallic glass model return to being nearly Arrhenius fashion at low $T$. Notably, quasi-thermodynamic measurements, such as specific heat measurements, show little or no evidence of any thermodynamic "feature" near $T_g$ in fluids exhibiting FS glass-formation so we have a thermodynamic signature to point to in relation to identifying the glass transition.[72] A tendency of the structural relaxation time to become Arrhenius in the $T$-regime below $T_g$ in which the material remains in a liquid state is predicted by the string model of the dynamics of GF liquids, an extension [27, 73] of the Adam and Gibbs model [74] that directly considers cooperative motion in cooled liquids and a statistical mechanical theory of the string-like dynamic clusters seen in MD simulations of model GF liquids.

The generalized entropy theory of glass-formation [41, 67] also indicates that there are two distinct regimes of glass-formation, a high and low $T$ regime where a characteristic $T$ separating these regimes is termed the "crossover temperature", $T_c$. As in our previous work on the Cu-Zr metallic glasses having a different material composition, we identify this "crossover temperature" by the occurrence of sharp deviation of $\langle u^2 \rangle$ from a linear variation, indicating the onset of strongly anharmonic interparticle interactions. In ordinary GF liquids, $\tau_\alpha$ scales as a power-law, $\tau_\alpha \sim [(T - T_c) / T_c]^{-\gamma_c}$, over a limited $T$ range above $T_c$, and we check below if this phenomenology applies to a material exhibiting FS glass-formation. The common observation of this type of power-law scaling in water would suggest that this scaling "feature" is general in GF liquids [75-76], but as in the VFT equation, the $T$-range where this scaling relation holds is limited. Nonetheless, the determination of the characteristic temperatures $T_o$ and $T_c$ are important temperatures even if no actual structural relaxation time divergence occurs at either of these $T$.

Finally, we consider the estimate of the "onset temperature" $T_A$ of glass-formation at which non-Arrhenius dynamics and other deviations from simple fluid dynamics emerge in the liquid dynamics. In our previous work on Cu-Zr metallic glasses, we found that we could obtain a good rough estimate of $T_A$ based on a Lindemann-type of criterion introduced by La Violette and Stillinger [67, 77] for the instability of the liquid state to local ordering. Following their work, we estimate $T_A$ from the condition that $\langle u^2 \rangle^{1/2} / \sigma$ is about 9 times its value at $T_g$. Dudowicz et al.[67] found that this simple criterion was remarkably consistent with $T_g$ estimates calculated through direct computation in the generalized entropy theory of glass-formation even in the case of polymer materials having different molecular architectures. Based on this simple criterion, we estimate $T_{A,u}$ where the "$u$" in the subscript that our estimate is based on $\langle u^2 \rangle$. In our previous simulations studies of the Al-Sm GF liquid [1-2], we estimated $T_A$ and $T_c$ by the traditional method of finding the $T$ at which $\tau_\alpha$ departs from being Arrhenius (See Section B of the Supplementary Material of Ref. 1) and the power-law scaling near the crossover temperature $T_c$ noted above. Notably, the characteristic temperature $T_c$ is precisely defined as the inflection point temperature of the product of the configurational entropy $S_c$ (i.e., non-vibrational entropy of the fluid), [67] times $T$, a quantity that can be calculated with precision from this molecular model of the thermodynamics of liquids. Based on the data summarized in Fig. S5 of Ref. 1, we estimated $T_A = 927$ K from a consideration of the $T$ dependence of the apparent activation energy as a function of $T$, which is reasonably consistent with the rough estimate of $T_{A,u} = 906$ K based on the rough $\langle u^2 \rangle$-based criterion indicated above. Our previous study of Cu-Zr metallic GF liquids having a wide range of composition also exhibited remarkably close correspondence between estimates of the characteristic temperatures of glass-formation from $\langle u^2 \rangle$ data with independently estimated values determined from structural relaxation time and mass diffusion data.

The FS transition in GF liquids signals some additional features that are not observed or conspicuous in OGF liquids. In particular, the local slope on the Arrhenius plot of the structural relaxation time or diffusion coefficient is often identified as being the activation energy and we denote this generally $T$-dependent quantity as $E_{diff}$. (It should be appreciated that $E_{diff}$ can be greatly different from the actual activation energy and this matter is discussed in Ref. 1.). In particular, the sharp increase in $E_{diff}$ first increases upon lowering $T$ below $T_A$, but this slope peaks, and then falls as the system is cooled further towards the glass state where Arrhenius relaxation and diffusion re-emerge (See Fig. 3 for an illustration of the observed behavior of $E_{diff}$ in our AL-Sm GF liquid.). The FS transition is evidently a transition between a high and low $T$ Arrhenius regimes, each having its own distinct activation energies and it is natural to define the intermediate $T$ at which the $E_{diff}$ peaks, the "FS transition temperature", $T_{FS}$. The maximum in $E_{diff}$ in Figure 3 indicates that characteristic temperature equals, $T_{FS} = 700$ K, which is evidently intermediate between $T_c$ and $T_\lambda$. We also clearly observe that the initial increase of $E_{diff}(T)$ upon cooling nearly coincides with $T_A$ and that $E_{diff}(T)$ saturates to a nearly constant value near 500 K, a $T$ close to our estimate of $T_g$ above. Thus, $T_A$, $T_\lambda$, and $T_g$ demark the beginning, middle and end of the glass transition. As noted in the previous section, we may also identify a temperature $T_o$ at which $\langle u^2 \rangle$ extrapolates to 0. In OGF liquids, this characteristic temperature coincides normally with the VFT temperature, but no such identification in our FS GF liquid since the VFT equation no longer describes $\tau_\alpha$ over large $T$ range as in OGF liquids. Despite the extrapolation of $\langle u^2 \rangle$ involved in estimating this characteristic temperature, we may anticipate that $T_o$ might correspond to the onset of solidification of our metallic glass into a "glass" state, regardless of the value of the fluid configurational entropy at $T_o$. We observe in the next section, as in previous simulations of polymeric GF liquids [41, 78], that the extrapolated value of $H$ as $T$ approaches $T_o$ nearly equals value on the order of $10^{-3}$, the critical value of *H* defining the onset of "effective hyperuniformity". This finding is apparently consistent with Torquato's concept of a "perfect glass" [79], defined as being an *equilibrium solid state* rather than just a non-equilibrium "frozen" liquid. This definition is generally different from the notion of an ideal non-equilibrium glass state hypothesized to exist by Adam and Gibbs when the fluid configurational entropy fluid approaches 0.

It is commonly observed in GF liquids that an extrapolation of the Arrhenius curve describing the Johari-Goldstein relaxation time $\tau_{JG}$ intersects the curve describing the α-relaxation time $\tau_\alpha$ at a *T* near the "crossover temperature" $T_c$. [80-81] We estimated this α-β "bifurcation temperature" $T_{\alpha\beta}$ from independent estimates of $\tau_{JG}$ and $\tau_\alpha$, and found that $T_{\alpha\beta}$ in the Al-Sm material.[1,2] is indeed close to $T_c$ so that this common, but not a universal feature of GF liquids, is apparently preserved in our Al-Sm GF liquid exhibiting FS glass-formation. This temperature is designated $T_{\alpha\beta}$ for comparison to the other characteristic temperatures in Fig. 3.

We also emphasize that FS GF liquids normally exhibit some rather distinctive features from OGF liquids that serve to define other characteristic temperatures of fluids undergoing FS glass formation. These fluids exhibit true thermodynamic "anomalies" that are not conspicuous in OGF liquids. For example, simulations of water have also indicated that the specific heat $C_p$ and isothermal compressibility, [82-83] exhibit a maximum at a common characteristic temperature, which can be taken as a definition of $T_\lambda$ and the inset of Fig. 3 shows that this same pattern of behavior arises in our Sm-Al metallic glass, as expected, where we see that $C_p$ and the thermal expansion coefficient both exhibit an extremum near *T* = 750 K. We also examined the 4-point density correlation function $\chi_4$ as a function and the noise exponent governing potential energy fluctuations, which likewise have a peak near this characteristic temperature, as seen before in simulations of water. [82-83] Our simulation estimates of these "linear response" properties exhibit remarkably similar trends to the corresponding properties of water, except for the important matter that the position of the peak in $C_p$ arises near $T_c$ in water, while $T_\lambda$ occurs well above $T_c$ in our Sm-Al metallic GF material. These characteristic temperatures are also included for comparison in Fig. 3. Now that we have determined the characteristic temperatures of glass-formation and briefly explained the basic phenomenology of FS glass-formation, we return to our discussion of the $T$ dependence of $H$.

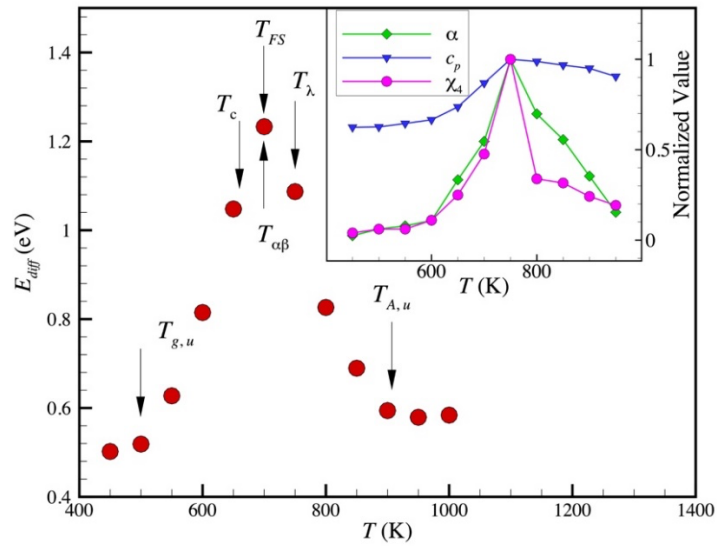

**Figure 3.** Differential activation energy $E_{\text{diff}}$ obtained from the diffusion coefficient of an $Al_{90}Sm_{10}$ metallic GF material with all characteristic temperatures labelled (See Ref. 1 for the estimation of this data). The inset shows specific heat, peak value of 4-point density function $\chi_4$ and the colored exponent α describing the power-law scaling of potential energy fluctuations with frequency as a function of $T$. We define the $T$ at which this specific maximum occurs to be the 'lambda temperature', $T_\lambda$. Compare the $T$ dependence of the response functions of our Al-Sm metallic glass-forming systems and corresponding computational estimates of these response functions for simulated water (e.g., see Fig. 3c of Ref. [83]).

The universal use of these critical "jamming" conditions to estimate the characteristic temperatures of glass-formation remains to be established for other fluids, but we hope that this type of condition proves to be practically useful, as in the case of the phenomenological Lindemann criterion [56, 62-63] and Hansen-Verlet criterion for freezing. [37-40]

**D. Characterization of the *T* dependence of *H***

We next utilize the information that we have established for the characteristic temperatures of glass-formation to interpret and quantify the *T* variation of *H* shown in Figs. 1 and 2. It is apparent from the enlargement of the *H* data in Fig. 2 that *H* varies more strongly than linearly with *T* and that this quantity does not extrapolate to 0 near $T_o$, as found in the case of $\langle u^2 \rangle$. Visual inspection of the data suggested to us that *H* might exhibit a power-law variation with *T*, and based on this heuristic argument, we fit our *H* data to the functional form,

$$H(T) \approx H_o + H_1 \left[(T - T_o) / T_o\right]^{\delta} \qquad (3)$$

where $H_o = 1.0 \times 10^{-3}$, $H_1 = 3.5 \times 10^{-4}$ and $\delta = 2.13$. The solid line in Fig. 2 shows a comparison of Eq. (3) to our admittedly somewhat scattered $H(T)$ data. We anticipated this non-linear scaling from the expected scaling between the isothermal compressibility, $\kappa_T \sim \langle u^2 \rangle^{3/2}$ under conditions when $S_p$ is nearly independent of *T*, i.e., low *T* where $\langle u^2 \rangle$ varies linearly with *T*. The fitted exponent 2.13 is larger than the expected value of 1.5 from the scaling relation between $\kappa_T$ and $\langle u^2 \rangle$, but the observed trend in *H* as *T* is varied qualitatively accords expectations.

Perhaps the most interesting outcome of this fitting procedure is the finding that *H* approaches $H_o = 1 \times 10^{-3}$ near $T_o$, the "critical value" of *H* defining "effective hyperuniformity". We thus find an indication that the critical *H* condition corresponds to an onset condition for amorphous solidification, provided that crystallization does not pre-empt this transition. Notably, the glass transition temperature $T_g$ appears to correspond to a significantly larger value of $H$ around 0.002 and the onset condition for non-Arrhenius dynamics seems to correspond to a much higher $H$ value near 0.007. Materials in the $T$ range between $T_o$ and $T_g$ are evidently still in a "viscous" fluid state in which physical aging effects are prevalent because of the large values of $\tau_\alpha$ in this $T$ range. The rheological state for $T$ below $T_o$ is contingent on whether can exist in an equilibrium thermodynamic state or not, which we conjecture is material-specific (see Conclusions).

These observations and the potential commonality of information contained in $\langle u^2 \rangle$ and $H$ then made us wonder about the magnitude of $H$ in crystalline materials at finite temperatures. It is well-known that $\langle u^2 \rangle$ increases with $T$ in crystalline materials where this quantity reaches a critical Lindemann value (somewhat dependent on the crystal symmetry and interaction type, but generally on the order $\langle u^2 \rangle^{1/2} / \sigma \approx 0.1$). Correspondingly, we might expect $H$ in crystals to likewise progressively increases with increasing $T$ until it reaches some "critical value" $H_c$ at which the crystal melts, where this critical $H$ value might even be equal to the corresponding value for "effective hyperuniformity" in cooled liquids signaling when they transform into solids. Of course, this somewhat naïve line of thinking contradicts blanket statements often made in the scientific literature to the effect that crystals are examples of perfectly hyperuniform materials for which $H = 0$. [6, 79] In a recent paper, Kim and Torquato examined the effect of imperfections, such as vacancies, interstitials, stochastic lattice displacement, and thermal excitations, on hyperuniformity in crystal systems , where it is indicated that such imperfections could degrade or destroy strict hyperuniformity for which $H = 0$. [84] This work did not make quantitative estimates of H for heated and defective crystals, however, so that the magnitude of $H$ in real materials has been uncertain before the present work.

Based on our current observations of $H$ in our Al-Sm metallic glass, we began to strongly suspect perfect hyperuniformity should not exist in any real matter, either crystalline or non-crystalline, at finite temperatures. Accordingly, we simulated a model crystalline Cu material that we have studied previously [85-86] over a range of $T$ to determine $H$ for this model material. As anticipated, we found that $H$ is indeed positive at finite $T$ and that this quantity increases progressively as the material is heated, just as one would expect from the qualitative relationship between $H$ and $\langle u^2 \rangle$ discussed above for our metallic GF liquid.

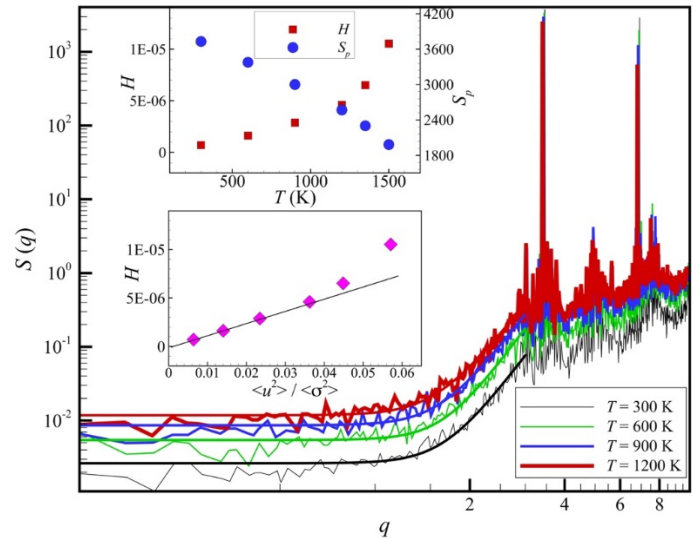

**Figure 4**. Static structure factor $S(q)$ of crystalline Cu over a range of temperatures below the melting temperature. The upper inset shows $H$, and the first peak height $S_p$ of $S(q)$ over a range of $T$ and $H$ as a function of $\langle u^2 \rangle / \langle \sigma^2 \rangle$ is shown as an inset to this lower figure inset where these quantities are found to be nearly linearly related in the low-temperature region. Although $H$ is significantly smaller in the crystalline material than in the glass-forming liquid, mainly due to the large value of $S_p$, the hyperuniformity index $H$ is certainly not zero in this model crystalline material as often stated. [6, 79]

Hyperuniformity is evidently a conceptual Platonic form [87] that can only be approached, but never physically reached in *equilibrium* materials at finite $T$. Moreover, the existence of a "critical" value of $H$ on the order 0.001 simply demarks a change of material condition in which long-wavelength density fluctuations are strongly suppressed due to the emergence of a "solid" material state. The physical significance of this state is that density fluctuations become energetically extremely costly so that these fluctuations become strongly suppressed in a $T$ or density range in which $H \leq 0.001$. In colloquial terminology, $H$ then characterizes the degree to which the material has become "jammed". We next turn to a consideration of the physical origin of the exceptionally low values of $H$ in liquids exhibiting fragile to strong glass-formation.

**E. Topological Correlations in Cooled Liquids and Emergent Hyperuniformity**

The observation of exceptionally low values of $H$ in materials undergoing FS glass-formation prompts the question of what physical attribute of cooled, typically network-forming liquids, should give rise to a general prevalence of this phenomenon in this class of glass-forming liquids. Douglas and coworkers [88] have previously argued that a thermodynamic polymerization transition (not strictly a phase transition because of variable transition rounding [89]) generally underlies glass-formation and we adopt this viewpoint as the basis of our following discussion.

While the formation of polymeric structures in cooled liquids seems to be a generic property of GF liquids, the topological form of the polymeric structures is evidently variable. The highly cooperative nature of FS GF suggests that the high cooperativity is linked to the randomly branched polymer structures that characteristic form in this type of GF liquid (Explaining the term "network glass-formers"). FS GF materials exhibit striking "anomalies" in their thermodynamic response functions[89-93], such as the specific heat, thermal expansion coefficient of the fluid, and isothermal compressibility that also accompany sharp changes in the viscosity and viscoelasticity of solutions exhibiting supramolecular self-assembly. Thermodynamic "anomalies" of this kind are particularly well-known in the case of water [44, 83], where this phenomenon is often attributed to an initially unanticipated liquid-liquid phase separation of water from itself [94]. Liquid-liquid phase separation is thought to arise from water's capacity to exhibit a multiplicity of distinct packing states having distinct thermodynamic and dynamic signatures - mobile high-density water and low-density immobile water. [44] Recent work by Sciortino and coworkers have suggested that the coexisting phases of water can be precisely characterized in terms of branched polymer phases having distinct topological structures. [95] This model of liquid-liquid phase separation in water would naturally explain the Holten and Anisimov model [96] of cooled water in terms of coexisting liquid phases having different entropies as polymer topology greatly influences polymer entropy. In previous work, we have seen the same anomalies in the thermodynamic response functions in our Al-Sm metallic glass material, and we strongly suspect that these anomalies arise from the topological correlations that arise in these materials at low temperatures. [1-2]   It is then of interest to consider what physical factors influence the cooperativity and topological nature of thermodynamic polymerization transitions.

Sharply defined features in thermodynamic response functions, such as the specific heat, density and osmotic compressibility are characteristic of equilibrium polymerization transitions that are *highly cooperative*. [91] The equilibrium polymerization of sulfur provides a particular example of equilibrium polymerization of an atomic fluid in which the transition is so cooperative that it greatly resembles a second-order phase transition, as evidenced by a sharp lambda transition in the specific heat $C_p$. [90] The cooperativity of activated and chemically initiated polymerization transitions [89, 92] can be *tuned* over a wide range by varying the initiator concentration, rate of activation or other physical processes that initiates or inhibits the polymerization process. [93] The degree of cooperativity of this class of transitions is directly related to the extent that the assembly process resembles a phase transition. [140] Evidently, "cooperativity" has a similar meaning in the fields of glass-formation and self-assembly processes, providing a novel perspective on the origin of fragility changes in glass-forming liquids. We next consider how this perspective might relate to understanding the fundamental origin of both "ordinary" and fragile-strong glass-formation. "Ordinary" glass-formation appears to be consistent with a highly "rounded" thermodynamic transition in which there is no observable or only a weak [97-98] thermodynamic signature of a "liquid-liquid" transition, while $C_p$ shows a large drop after the initial rising in cooled liquids as the material goes out of equilibrium, a purely kinetic phenomenon that is often taken as the *definition* of $T_g$.

There has long been discussion and controversy relating to a putative "liquid-liquid transition temperature" $T_{LL}$ in polymer melts and other ordinary GF liquids, [98-108] where $T_{LL}$ has typically been reported to be in the range, $T_{LL} \approx (1.2 \text{ to } 1.3) T_g$.[99-106, 109] Flory and coworkers [97, 110] claimed to have observed a third-order phase transition at a $T$ well above $T_g$ in polystyrene. However, the lack of any theoretical rationale and the subtle nature of the observed thermodynamic signatures defining this transition has made even the existence of such a transition temperature controversial in the academic material science community, until recently. [98-108] However, the practical importance of $T_{LL}$ is broadly recognized in the field of process engineering because this $T$ often signals gross changes in fluid flow and diffusion processes that are highly relevant for material processing. [99-106] In contrast, GF fluids exhibiting FS GF exhibit a $C_p$ peak that follows a similar phenomenology as $T_{LL}$ in relation to its position relative to $T_g$, but the intensity of the thermodynamic feature, and the occurrence of anomalies in other thermodynamic response functions, suggest that this characteristic temperature corresponds to some genuine type of thermodynamic transition in the material. [1, 43-44]

We interpret these observations as implying that there are possible unifying features in GF fluids exhibiting both ordinary and FS glass-formation. The challenge is then to develop a description of GF liquids that encompasses both classes of GF materials. Within the equilibrium polymerization model of glass-formation described above based on the formation of linear polymer chains, materials exhibiting FS glass-formation should correspond to a highly cooperative variety of polymerization transition if both ordinary and FS types of glass-formation can be described within a unified model of this kind.

So, what aspect of fragile-strong glass-forming liquids explain the high cooperativity of this type of glass-formation? We suggest that the *multifunctional nature* of the associations that give rise to load-bearing branched polymer structures in these polymer materials is responsible for the cooperativity of glass-formation process in this class of materials which is often classified as corresponding to "network-forming" glass materials. Clear evidence for high cooperativity in branched equilibrium polymerization can be seen in the emergence of *multiple critical points* in single component fluids exhibiting branched equilibrium polymerization.[111] High cooperativity in linear chain equilibrium polymerization, on the other hand, requires the constraint of a low rate of activation or a small amount of initiator.[91] The propensity of fluids exhibiting FS GF to form network structures upon cooling in the glass material might then be sufficient to explain the occurrence of both fragile-strong glass-formation, while the formation dynamic linear polymer structures might account correspondingly for "ordinary" GF liquids. Douglas and Hubbard [112] previously suggested that the occurrence of topologically distinct equilibrium polymerization processes in different classes of GF liquids should lead to distinct signatures in the stress relaxation of network forming and ordinary GF liquids.

Our hypothesis that the fragile-strong glass-formation is a consequence of the branched nature of the dynamic polymerization process underlying glass-formation also has significant consequences for emergent hyperuniformity at low temperatures in this class of GF liquids. The packing correlations derived from branching when the polymers chains within the branched polymer are not very long tend to lead to materials that are denser than their linear chain counterparts. [113-114] Moreover, simulations of branching of polymer branching in the form of model star and bottlebrush polymer have indicated a significant reduction of $H$ values for these polymer structures relative to linear polymers in the melt state, which in turn have lower $H$ values than the non-polymeric monomers from which they are composed. Polymer branching by itself can evidently lead to an approach to effective hyperuniformity. The specific form of topology and the form of loops and other topological constraints can be expected to modulate these topological correlations, as investigated in some detail in the case of knotting in the case of the average dimensions of isolated knotted ring polymers in solution [115-116] and the density of knotted ring polymers in the melt [117], and in the effect of branching density on the dimensions of isolated randomly branched or "network" polymers. [118-119] It has also been observed that when the topological complexity of branched polymer structures, defined appropriately for the particular type of branched polymer under discussion, e.g., average crossing number in knotted ring polymers, star arms in star polymers, etc., becomes large, then the branched polymers become more rigid, and this topological rigidification [115-116] can lead to a diminished packing efficiency as defined by the density. There are thus competing effects arising from the topological structures of the self-assembled molecules that can influence the ultimate thermodynamic and dynamic properties of the material in general. Nonetheless, we may expect the formation of structures of modest topological complexity in cooled liquids to lead to a reduced relative value of $H,$ and we suggest that this effect is ultimately responsible for the relatively small values of *H* that we observe in our simulated model Al-Sm metallic GF liquid, and for the widely observed relatively small values of *H* observed in earlier simulations of network-forming GF liquids discussed above. This interpretation of the low *H* values in these liquids is also consistent with the propensity for liquid-liquid phase separation and a highly cooperative form of dynamic properties of glass-formation that are characteristic of fragile-strong GF, as well as by well-defined signatures in thermodynamic response functions of these materials that are not as "noticeable" in ordinary GF liquids. Importantly, this interpretation of fragile-strong GF provides a unified view with that of OGF, which explains why the tendency for the formation of dynamic polymeric structures as a common physical manifestation of dynamic heterogeneity in GF liquids [1] is quite understandable. We should then look to subtle differences in the topological form of these dynamic polymer structures that induce correlations in the material thermodynamic and dynamic properties.

Polymerization transitions are characterized by a drop of configurational entropy, and variable degrees of thermodynamic sharpness [89] as the temperature is varied through the transition temperature so it becomes necessary to delineate this type of broad ("rounded") thermodynamic transition by characterizing the points where the transition "begins", peaks in the middle, and "ends" rather than just a single phase transition temperature. [88, 112, 120-121] Even though there is normally no long-range translational order in this type of self-assembly transition, the reduction of the configurational entropy upon passing through this type of transition is indicative of a type of "ordering" process that derives from the correlated relation of the particles within the polymeric structures and the associated topological correlations on the dynamics and thermodynamics of these materials. In a broad sense, we may view both crystallization and quasi-crystal formation, and even protein folding, as being particular types of polymerization transitions that exhibit different types of "structural" correlations in their respective "ordered" states. We may thus arrive at a general theoretical framework that subsumes "solidification" by crystallization and glass-formation in which differences in these materials can be described by long-range correlations arising from positional, orientational and topological correlations deriving from the structure of the material.

The formation of dynamic clusters of atoms in cooled liquids can be expected to have significant ramifications for the scattering properties of GF liquids when the subunits of these dynamic polymeric structures ("equilibrium polymers") have a different density or other sources of scattering contrast with the surrounding fluid exhibiting a distinct local structure. The polymeric nature of these clusters should then enable the modeling of scattering observations [27] on GF liquids to gain quantitative information about this type of supramolecular organization process involved, and this observation leads us to return to our discussion of the low-$q$ upturn and the pre-peak scattering features apparent in Fig. 1, features that are rather common in GF liquids. [52] This is evidently a problem of wide scope and deserving of a separate publication devoted to this topic, but in the Supplementary Material section, we point out some available models and simulation observations that should be helpful in analyzing this type of scattering data that are often ignored because of the lack of any accepted theoretical framework for their interpretation.

**IV. Conclusions**

Many recent studies of network-forming glass-forming liquids have indicated that these materials exhibit Fragile-Strong (FS) glass-formation corresponding to a qualitatively different phenomenology in their dynamics than "ordinary" glass-forming liquids, along with striking anomalies in the thermodynamic response functions of these liquids that are normally not apparent in other liquids. Moreover, liquid-liquid phase separation has commonly been reported for this class of materials, as well as a propensity to form hyperuniform materials with low hyperuniformity index $H$ values at low temperatures that are rarely observed in other glass-forming liquids under equilibrium or near equilibrium conditions. In recent work, we unexpectedly observed that the dynamical and thermodynamic properties of an Al-Sm metallic glass-forming material exhibited all the phenomenological hallmarks of FS glass-formation, which prompted us to attempt to better understand the physical origin of this type of glass-formation. Motivated by previous studies indicating an apparently general tendency of FS glass-formers to exhibit low $H$ values, a quantitative measure of molecular "jamming", and network formation of the substituent molecules upon approaching the glass-transition, we estimated $H$ for our simulated Al-Sm metallic glass [1] and found that this "packing parameter" was indeed exceptionally small in comparison to other simulated cooled liquids that we have studied previously under equilibrium conditions. Since the origin of an often stated critical value of $H$ on the order of $10^{-3}$ for the emergence of an "effectively hyperuniform" state seemed obscure to us, we determined the temperature dependence of $H$ along with another more familiar "jamming parameter, the Debye-Waller parameter $\langle u^2 \rangle$, which is measurable by a variety of experimental techniques and which is often used to estimate the melting temperature of crystalline materials and the glass-transition in temperature based on well-known phenomenological Lindemann criteria. Further, since the Lindemann criterion is more commonly considered in crystalline materials, we also considered a parallel analysis of the $T$ dependence of $H$ and $\langle u^2 \rangle$ for a model crystalline Cu material. This comparative analysis proved that the information content of $H$ is closely related to that of $\langle u^2 \rangle$ and that the often-stated critical $H$ index value, $H_c = 10^{-3}$, can be understood as an order of magnitude condition for amorphous solidification that can be defined in the same spirit as the empirical Lindemann criterion. In qualitative terms, this condition describes the physical condition in which the compressibility of the material is reduced to such a critical degree that the material exhibits solid-like rather than liquid-like characteristics, even if the liquid has no long-range positional or orientational order. Of course, both crystalline and quasi-crystalline materials generally meet this "effective hyperuniformity" condition and can be naturally classified as being "solids" by this criterion. One new outcome of these calculations was the finding that $H$ can be appreciable in equilibrium crystalline materials at finite temperature, tending towards the value of liquids as the melting temperature $T_m$ from below, paralleling the increase of $\langle u^2 \rangle$ upon heating towards $T_m$, which is a contrast to common statements in the scientific literature that $H = 0$ in both crystalline and quasi-crystalline materials. Exact hyperuniformity ($H = 0$) under equilibrium conditions only arises at the limit of $T = 0$.

Although the clarification of the physical meaning of the $H$ in materials in equilibrium is one of the important contributions of the present work, our analysis of $H$ in our model metallic glass also provides insight into the relatively low $H$ values seen in glass-forming liquids exhibiting FS glass-formation. We discuss evidence indicating that low values of $H$ derive from the topological correlations in the fluid that in turn derive from the tendency of the molecules to form supramolecular polymer structures in cooled liquids, structures that form the structural basis of "dynamic heterogeneity" in cooled liquids. Based on this perspective, glass-formers that exhibit FS glass-formation are identified as "network-forming" glass-formers in which the higher topological complexity of the self-assembly process induces topological correlations that can greatly influence both the thermodynamic and dynamic properties of the glass-forming material. This point of view of FS glass-formation also explains the high cooperativity of this type of glass-formation and the propensity for liquid-liquid phase separation in these materials, even when the material is comprised of a single molecular species. We thus arrive at a potentially unified framework for understanding both FS and "ordinary" GF liquids. However, further work will be required to quantify the topological structures of glass-forming liquids and the nature of the correlations they induce.

There are also practical implications of the approach to non-uniformity for simulation studies of glass-formation. The approach of liquids to a state of effective hyperuniformity at low temperatures has been predicted to imply that the direct correlation function, which can be approximated by the potential of the mean interaction between the particles, develops long-range *correlations* of a similar mathematical form to the pair correlation function approaching a liquid-vapor critical point so that hyperuniformity is in a sense the antithesis of ordinary critical fluid behavior. [4, 41] These long-range correlations can be expected [40] to give rise to appreciable finite-size effects, which should be a general matter of concern for simulations of GF liquids since $H$ has been observed to approach hyperuniformity conditions even in simulations of ordinary polymeric GF liquids. [4, 41] Torquato and coworkers found that the well-known sum rule relating $S(0)$ to the density and isothermal compressibility starts to become violated in model GF liquids even at relatively high $T$. [48, 122] One interpretation of these disturbing observations is that simulations in the $T$ range of greatest interest for applications are inherently out of equilibrium! While this effect is quite real, we alternatively tentatively interpret this apparent deviation from this fundamental thermodynamic relation to arise from finite-size effects upon approaching an effectively hyperuniform glass state. Indeed, a simulation study by Sastry and coworkers [123] has provided clear evidence of appreciable finite size effects in a model GF liquid where the characteristic scale derived from the standard finite-size scaling analysis coincides within numerical uncertainty with the characteristic scale $\xi_4$ associated with the 4-point density correlation function, $\chi_4$. This correlation function [27] heavily weights the immobile particle clusters so that this growing characteristic scale of GF liquids is almost certainly related to the size of the immobile particles in the cooled liquids. It seems possible that $\xi_4$ might be related to a corresponding growing scale derived from the fluid direct correlation function that underlies Torquato's theory of amorphous solidification.[33] Torquato's theory of glass solidification is built around the concept of emergent hyperuniformity in liquids that have been cooled to a sufficiently $T$ to transform into an equilibrium solid, while maintaining thermodynamic equilibrium. Establishing this type of linkage would help create a theoretical foundation for describing glass formation. Apart from matters of fundamental interest, the development of these long-range correlations and associated finite-size effects, even at much higher $T$ than those in which the fluid is effectively [41] hyperuniform, adds to the difficulty of the growing relaxation times and slowing of diffusion in simulating the properties of GF liquids. These finite-size effects evidently require further investigation.

The higher cooperativity of the FS glass-formation also has practical implications for the observability of a thermodynamically defined transition in the material. Upon approaching the glass transition in ordinary GF liquids by progressively cooling, the structural relaxation time grows so large that the system cannot "complete" the thermodynamic transition before the material seizes up through a non-equilibrium structural arrest, leading to "features" in quasi-thermodynamic measurements such as the specific heat, density, etc. that purely reflect the fact that the material has gone out of equilibrium. The relatively rapid rate at which systems exhibiting a FS type glass-formation "complete" the thermodynamic transition allows it to exhibit well-defined thermodynamic features characteristic of materials exhibiting a self-assembly transition as well as an observable low-temperature Arrhenius dynamics regime in some cases. Corresponding changes in the variation of the rate of change of the configurational entropy of these classes of glass-forming fluids when *T* is varied would seem to account for whether the glass-transition is signalled by a non-equilibrium or equilibrium peak in the specific heat and other thermodynamic properties for "ordinary" and FS glass-forming liquids, respectively.

Phase-change memory materials provide an important class of applications in which FS glass-formation is apparently important for practical device performance. These materials are often compounded from Te, and other chalcogenides and group-IV and group-V elements [124] known to exhibit equilibrium polymerization in the liquid state upon cooling (See Supplementary Information) along with the FS nature of glass-formation in these materials allows for rapid switching between crystalline and amorphous states, the states in which information is stored. The rapidity of this switching of material states enhances the speed of data recording, and the ultra-stable nature of this class of materials in their glass state [125-126] aids in the stability or "non-volatility" of the stored information based on these materials. [127-128] Large changes in conductivity and other properties also accompany this transition which can be beneficial in the applications of these materials. In a metallic glass context, FS GF is sometimes accompanied by a significant increase ($\approx$ 20 %) in the material hardness[126, 129], which we suggest is a natural consequence of emergent hyperuniformity. We may expect many further applications of FS glass-formation in the future because of the relatively rapid rate at which the glass formation can be actuated and the occurrence of properties in the glass state that are reminiscent of crystalline materials, which are likewise materials having small values of *H*.

## Acknowledgements

H.Z. and X.Y.W. gratefully acknowledge the support of the Natural Sciences and Engineering Research Council of Canada under the Discovery Grant Program (RGPIN-2017-03814) and Accelerator Supplements (RGPAS-2017- 507975).

## Data Availability Statement

The data that supports the findings of this study are available within the article and its supplementary material section.

## Supplementary Information

See this section for specialized topics related to the main body of the paper: These sections have the following titles that define the topics considered: 1) Fragile-Strong Transition, Liquid-liquid Transition, and Equilibrium Polymerization, 2) Specific Examples of Fluids Exhibiting Equilibrium Polymerization, 3) Equilibrium Polymerization on Glass-Formation in Multiple Component Materials, 4) Suggested Physical Origin of the Widom Line, 4) Relation Between FS Glass-Formers and Superionic Crystalline Materials?

## Author Contribution Statement

H.Z., H.B.Y. and J.F.D. developed the idea and designed the simulation. X.Y.W., H.B.Y., and J.R.Z. performed MD simulations. H.Z., X.Y.W., J.R.Z. and J.F.D. analyzed the data. H.Z. and J.F.D. wrote the manuscript.

# References

1. Zhang, H.; Wang, X. Y.; Yu, H. B.; Douglas, J. F., Dynamic Heterogeneity, Cooperative Motion, and Johari-Goldstein Beta-Relaxation in a Metallic Glass-Forming Material Exhibiting a Fragile-to-Strong Transition. *Eur Phys J E* **2021**, *44*.
2. Zhang, H.; Wang, X. Y.; Yu, H. B.; Douglas, J. F., Fast Dynamics in a Model Metallic Glass-Forming Material *Journal of Chemical Physics* **2021**, *154*, 084505.
3. Zhang, H.; Zhong, C.; Douglas, J. F.; Wang, X. D.; Cao, Q. P.; Zhang, D. X.; Jiang, J. Z., Role of String-Like Collective Atomic Motion on Diffusion and Structural Relaxation in Glass Forming Cu-Zr Alloys. *Journal of Chemical Physics* **2015**, *142*, 164506.
4. Torquato, S., Hyperuniform States of Matter. *Phys Rep* **2018**, *745*, 1-95.
5. Martelli, F.; Torquato, S.; Giovambattista, N.; Car, R., Large-Scale Structure and Hyperuniformity of Amorphous Ices. *Phys. Rev. Lett.* **2017**, *119*, 136002.
6. Martelli, F., Steady-Like Topology of the Dynamical Hydrogen Bond Network in Supercooled Water. *PNAS Nexus* **2022**, *1*, 1-8.
7. Xie, R. B.; Long, G. G.; Weigand, S. J.; Moss, S. C.; Carvalho, T.; Roorda, S.; Hejna, M.; Torquato, S.; Steinhardt, P. J., Hyperuniformity in Amorphous Silicon Based on the Measurement of the Infinite-Wavelength Limit of the Structure Factor. *P Natl Acad Sci USA* **2013**, *110*, 13250-13254.
8. Dahal, D.; Atta-Fynn, R.; Elliott, S. R.; Biswas, P., Hyperuniformity and Static Structure Factor of Amorphous Silicon in the Infinite-Wavelength Limit. In *J Phys Conf Ser*, Landau, D. P.; Bachmann, M.; Lewis, S. P.; Schuttler, H. B., Eds. Iop Publishing Ltd: Bristol, 2019; Vol. 1252, p 012003.
9. Zheng, Y., et al., Disordered Hyperuniformity in Two-Dimensional Amorphous Silica. *Sci. Adv.* **2020**, *6*, 5.
10. Chremos, A.; Douglas, J. F., Hidden Hyperuniformity in Soft Polymeric Materials. *Phys. Rev. Lett.* **2018**, *121*, 258002.
11. Chremos, A., Design of Nearly Perfect Hyperuniform Polymeric Materials. *Journal of Chemical Physics* **2020**, *153*, 054902.
12. Akcora, P., et al., Anisotropic Self-Assembly of Spherical Polymer-Grafted Nanoparticles. *Nat Mater* **2009**, *8*, 354-U121.
13. Kannemann, F. O.; Becke, A. D., Atomic Volumes and Polarizabilities in Density-Functional Theory. *Journal of Chemical Physics* **2012**, *136*, 034109.
14. Saika-Voivod, I.; Smallenburg, F.; Sciortino, F., Understanding Tetrahedral Liquids through Patchy Colloids. *Journal of Chemical Physics* **2013**, *139*, 234901.
15. Hsu, C. W.; Largo, J.; Sciortino, F.; Starr, F. W., Hierarchies of Networked Phases Induced by Multiple Liquid-Liquid Critical Points. *P Natl Acad Sci USA* **2008**, *105*, 13711-13715.
16. Dai, W.; Hsu, C. W.; Sciortino, F.; Starr, F. W., Valency Dependence of Polymorphism and Polyamorphism in DNA-Functionalized Nanoparticles. *Langmuir* **2010**, *26*, 3601-3608.
17. Starr, F. W.; Sciortino, F., "Crystal-Clear" Liquid-Liquid Transition in a Tetrahedral Fluid. *Soft Matter* **2014**, *10*, 9413-9422.
18. Smallenburg, F.; Filion, L.; Sciortino, F., Erasing No-Man's Land by Thermodynamically Stabilizing the Liquid-Liquid Transition in Tetrahedral Particles. *Nature Physics, in press* **2014**, *10*, 653-657.
19. Mendelev, M. I.; Zhang, F.; Ye, Z.; Sun, Y.; Nguyen, M. C.; Wilson, S. R.; Wang, C. Z.; Ho, K. M., Development of Interatomic Potentials Appropriate for Simulation of Devitrification of Al90sm10 Alloy. *Model Simul Mater Sc* **2015**, *23*, 11.
20. Finnis, M. W.; Sinclair, J. E., A Simple Empirical N-Body Potential for Transition-Metals. *Philos Mag A* **1984**, *50*, 45-55.
21. Sun, Y.; Zhang, Y.; Zhang, F.; Ye, Z.; Ding, Z. J.; Wang, C. Z.; Ho, K. M., Cooling Rate Dependence of Structural Order in Al90sm10 Metallic Glass. *J. Appl. Phys.* **2016**, *120*, 015901.

22. Parrinello, M.; Rahman, A., Polymorphic Transitions in Single-Crystals - a New Molecular-Dynamics Method. *J. Appl. Phys.* **1981**, *52*, 7182-7190.
23. Nose, S., A Unified Formulation of the Constant Temperature Molecular-Dynamics Methods. *Journal of Chemical Physics* **1984**, *81*, 511-519.
24. Hoover, W. G., Canonical Dynamics - Equilibrium Phase-Space Distributions. *Physical Review A* **1985**, *31*, 1695-1697.
25. Plimpton, S., Fast Parallel Algorithms for Short-Range Molecular-Dynamics. *Journal of Computational Physics* **1995**, *117*, 1-19.
26. Mishin, Y.; Mehl, M. J.; Papaconstantopoulos, D. A.; Voter, A. F.; Kress, J. D., Structural Stability and Lattice Defects in Copper: Ab Initio, Tight-Binding, and Embedded-Atom Calculations. *Phys. Rev. B* **2001**, *63*, 224106.
27. Starr, F. W.; Douglas, J. F.; Sastry, S., The Relationship of Dynamical Heterogeneity to the Adam-Gibbs and Random First-Order Transition Theories of Glass Formation. *Journal of Chemical Physics* **2013**, *138*, 12A541.
28. Wang, X. Y.; Xu, W. S.; Zhang, H.; Douglas, J. F., Universal Nature of Dynamic Heterogeneity in Glass-Forming Liquids: A Comparative Study of Metallic and Polymeric Glass-Forming Liquids. *Journal of Chemical Physics* **2019**, *151*, 184503.
29. Sun, Y.; Zhang, Y.; Zhang, F.; Ye, Z.; Ding, Z. J.; Wang, C. Z.; Ho, K. M., Cooling Rate Dependence of Structural Order in Al90sm10 Metallic Glass. *J. Appl. Phys.* **2016**, *120*, 6.
30. Cheng, Y. Q.; Ma, E., Atomic-Level Structure and Structure-Property Relationship in Metallic Glasses. *Progress in Materials Science* **2011**, *56*, 379-473.
31. Ding, J.; Ma, E., Computational Modeling Sheds Light on Structural Evolution in Metallic Glasses and Supercooled Liquids. *npj Comput. Mater.* **2017**, *3*, 12.
32. Torquato, S.; Stillinger, F. H., Local Density Fluctuations, Hyperuniformity, and Order Metrics. *Phys. Rev. E* **2003**, *68*, 041113.
33. Torquato, S., Perspective: Basic Understanding of Condensed Phases of Matter Via Packing Models. *Journal of Chemical Physics* **2018**, *149*, 020901.
34. Lomba, E.; Weis, J. J.; Torquato, S., Disordered Hyperuniformity in Two-Component Nonadditive Hard-Disk Plasmas. *Phys. Rev. E* **2017**, *96*, 062126.
35. Lomba, E.; Weis, J. J.; Torquato, S., Disordered Multihyperuniformity Derived from Binary Plasmas. *Phys. Rev. E* **2018**, *97*, 010102.
36. Chremos, A.; Douglas, J. F., Particle Localization and Hyperuniformity of Polymer-Grafted Nanoparticle Materials. *Ann Phys-Berlin* **2017**, *529*, 1600342.
37. Hansen, J. P.; Verlet, L., Phase Transitions of Lennard-Jones System. *Phys. Rev.* **1969**, *184*, 151-&.
38. Hansen, J. P., Phase Transition of Lennard-Jones System .2. High-Temperature Limit. *Physical Review A* **1970**, *2*, 221-&.
39. Hoffmann, G. P.; Lowen, H., Freezing and Melting Criteria in Non-Equilibrium. *Journal of Physics-Condensed Matter* **2001**, *13*, 9197-9206.
40. Costigliola, L.; Schroder, T. B.; Dyre, J. C., Freezing and Melting Line Invariants of the Lennard-Jones System. *Physical Chemistry Chemical Physics* **2016**, *18*, 14678-14690.
41. Xu, W. S.; Douglas, J. F.; Freed, K. F., Influence of Cohesive Energy on the Thermodynamic Properties of a Model Glass-Forming Polymer Melt. *Macromolecules* **2016**, *49*, 8341-8354.
42. Ito, K.; Moynihan, C. T.; Angell, C. A., Thermodynamic Determination of Fragility in Liquids and a Fragile-to-Strong Liquid Transition in Water. *Nature* **1999**, *398*, 492-495.
43. Shi, R.; Russo, J.; Tanaka, H., Origin of the Emergent Fragile-to-Strong Transition in Supercooled Water. *P Natl Acad Sci USA* **2018**, *115*, 9444-9449.
44. Shi, R.; Russo, J.; Tanaka, H., Common Microscopic Structural Origin for Water's Thermodynamic and Dynamic Anomalies. *Journal of Chemical Physics* **2018**, *149*, 224502.
45. Saika-Voivod, I.; Poole, P. H.; Sciortino, F., Fragile-to-Strong Transition and Polyamorphism in the Energy Landscape of Liquid Silica. *Nature* **2001**, *412*, 514-517.

46. Sun, Q. J.; Zhou, C.; Yue, Y. Z.; Hu, L. N., A Direct Link between the Fragile-to-Strong Transition and Relaxation in Supercooled Liquids. *J. Phys. Chem. Lett.* **2014**, *5*, 1170-1174.
47. Lucas, P., Fragile-to-Strong Transitions in Glass Forming Liquids. *Journal of Non-Crystalline Solids: X* **2019**, *4*, 100034.
48. Marcotte, E.; Stillinger, F. H.; Torquato, S., Nonequilibrium Static Growing Length Scales in Supercooled Liquids on Approaching the Glass Transition. *Journal of Chemical Physics* **2013**, *138*, 12A508.
49. Rapaport, D. C., *The Art of Molecular Dynamics Simulation*, 2nd ed.; Cambridge University Press: Cambridge, UK ; New York, NY, 2004, p xiii, 549 p.
50. Huang, C., et al., The Inhomogeneous Structure of Water at Ambient Conditions. *P Natl Acad Sci USA* **2009**, *106*, 15214-15218.
51. Zhuravlyov, V.; Goree, J.; Douglas, J. F.; Elvati, P.; Violi, A., Comparison of the Static Structure Factor at Long Wavelengths for a Dusty Plasma Liquid and Other Liquids. *Phys. Rev. E* **2022**, *Accepted*.
52. Fischer, E. W., Light-Scattering and Dielectric Studies on Glass-Forming Liquids. *Physica A* **1993**, *201*, 183-206.
53. Patkowski, A.; Thurn-Albrecht, T.; Banachowicz, E.; Steffen, W.; Bosecke, P.; Narayanan, T.; Fischer, E. W., Long-Range Density Fluctuations in Orthoterphenyl as Studied by Means of Ultrasmall-Angle X-Ray Scattering. *Phys. Rev. E* **2000**, *61*, 6909-6913.
54. Patkowski, A.; Fischer, E. W.; Steffen, W.; Glaser, H.; Baumann, M.; Ruths, T.; Meier, G., Unusual Features of Long-Range Density Fluctuations in Glass-Forming Organic Liquids: A Rayleigh and Rayleigh-Brillouin Light Scattering Study. *Phys. Rev. E* **2001**, *63*, 061503.
55. Zhou, C.; Hu, L. N.; Sun, Q. J.; Zheng, H. J.; Zhang, C. Z.; Yue, Y. Z., Structural Evolution During Fragile-to-Strong Transition in Cuzr(Al) Glass-Forming Liquids. *Journal of Chemical Physics* **2015**, *142*, 064508.
56. Douglas, J. F.; Xu, W. S., Equation of State and Entropy Theory Approach to Thermodynamic Scaling in Polymeric Glass-Forming Liquids. *Macromolecules* **2021**, *54*, 3247-3269.
57. Zhang, W. G.; Starr, F. W.; Douglas, J. F., Activation Free Energy Gradient Controls Interfacial Mobility Gradient in Thin Polymer Films. *Journal of Chemical Physics* **2021**, *155*, 174901.
58. Zheng, X. R.; Guo, Y. F.; Douglas, J. F.; Xia, W. J., Understanding the Role of Cross-Link Density in the Segmental Dynamics and Elastic Properties of Cross-Linked Thermosets. *Journal of Chemical Physics* **2022**, *157*, 064901.
59. Zheng, X. R.; Guo, Y. F.; Douglas, J. F.; Xia, W. J., Competing Effects of Cohesive Energy and Cross-Link Density on the Segmental Dynamics and Mechanical Properties of Cross-Linked Polymers. *Macromolecules* **2022**, *55*, 9990-10004.
60. Zhang, H.; Srolovitz, D. J.; Douglas, J. F.; Warren, J. A., Grain Boundaries Exhibit the Dynamics of Glass-Forming Liquids. *P Natl Acad Sci USA* **2009**, *106*, 7735-7740.
61. Douglas, J. F.; Betancourt, B. A. P.; Tong, X. H.; Zhang, H., Localization Model Description of Diffusion and Structural Relaxation in Glass-Forming Cu-Zr Alloys. *J Stat Mech-Theory E* **2016**, *2016*, 054048.
62. Gilvarry, J. J., Temperature-Dependent Equations of State of Solids. *J. Appl. Phys.* **1957**, *28*, 1253-1261.
63. Cartz, L., Thermal Vibrations of Atoms in Cubic Crystals .1. The Temperature Variation of Thermal Diffuse Scattering of X-Rays by Lead Single Crystals. *P Phys Soc Lond B* **1955**, *68*, 951-956.
64. Vogel, H., The Temperature Dependence Law of the Viscosity of Fluids. *Physikalische Zeitschrift* **1921**, *22*, 645-646.
65. Fulcher, G. S., Analysis of Recent Measurements of the Viscosity of Glasses. *Journal of the American Ceramic Society* **1925**, *8*, 339-355.
66. Tammann, G.; Hesse, W., The Dependancy of Viscosity on Temperature in Hypothermic Liquids. *Z. Anorg. Allg. Chem.* **1926**, *156*, 14.
67. Dudowicz, J.; Freed, K. F.; Douglas, J. F., *Advances in Chemical Physics. Vol. 137*; John Wiley & Sons, Inc.: New York, 2007; Vol. 137, p 245.

68. Xu, W. S.; Douglas, J. F.; Xia, W. J.; Xu, X. L., Investigation of the Temperature Dependence of Activation Volume in Glass-Forming Polymer Melts under Variable Pressure Conditions. *Macromolecules* **2020**, *53*, 6828-6841.
69. Miller, A. A., Glass-Transition Temperature of Water. *Science* **1969**, *163*, 1325-&.
70. Angell, C. A.; Sare, E. J.; Bressel, R. D., Concentrated Electrolyte Solution Transport Theory: Directly Measured Glass Temperatures and Vitreous Ice. *Journal of Physical Chemistry* **1967**, *71*, 2759-&.
71. Johari, G. P.; Hallbrucker, A.; Mayer, E., The Glass–Liquid Transition of Hyperquenched Water. *Nature* **1987**, *330*, 552-553.
72. Angell, C. A.; Shuppert, J.; Tucker, J. C., Nomalous Properties of Supercooled Water. Heat Capacity, Expansivity, and Proton Magnetic Resonance Chemical Shift from 0 to -38 C. *Journal of Physical Chemistry* **1973**, *77*, 3092-3099.
73. Xu, W. S.; Douglas, J. F.; Sun, Z. Y., Polymer Glass Formation: Role of Activation Free Energy, Configurational Entropy, and Collective Motion. *Macromolecules* **2021**, *54*, 3001-3033.
74. Adam, G.; Gibbs, J. H., On Temperature Dependence of Cooperative Relaxation Properties in Glass-Forming Liquids. *Journal of Chemical Physics* **1965**, *43*, 139-146.
75. Starr, F. W.; Sciortino, F.; Stanley, H. E., Dynamics of Simulated Water under Pressure. *Phys. Rev. E* **1999**, *60*, 6757-6768.
76. De Marzio, M.; Camisasca, G.; Rovere, M.; Gallo, P., Mode Coupling Theory and Fragile to Strong Transition in Supercooled Tip4p/2005 Water. *Journal of Chemical Physics* **2016**, *144*, 8.
77. Laviolette, R. A.; Stillinger, F. H., Multidimensional Geometric Aspects of the Solid Liquid Transition in Simple Substances. *Journal of Chemical Physics* **1985**, *83*, 4079-4085.
78. Xu, W. S.; Douglas, J. F.; Freed, K. F., Influence of Pressure on Glass Formation in a Simulated Polymer Melt. *Macromolecules* **2017**, *50*, 2585-2598.
79. Zhang, G.; Stillinger, F. H.; Torquato, S., The Perfect Glass Paradigm: Disordered Hyperuniform Glasses Down to Absolute Zero. *Sci Rep-Uk* **2016**, *6*, 36963.
80. Arbe, A.; Richter, D.; Colmenero, J.; Farago, B., Merging of the Alpha and Beta Relaxations in Polybutadiene: A Neutron Spin Echo and Dielectric Study. *Phys. Rev. E* **1996**, *54*, 3853-3869.
81. Rossler, E., Indications for a Change of Diffusion Mechanism in Supercooled Liquids. *Phys. Rev. Lett.* **1990**, *65*, 1595-1598.
82. Jana, B.; Bagchi, B., Intermittent Dynamics, Stochastic Resonance and Dynamical Heterogeneity in Supercooled Liquid Water. *J Phys Chem B* **2009**, *113*, 2221-2224.
83. Jana, B.; Singh, R. S.; Bagchi, B., String-Like Propagation of the 5-Coordinated Defect State in Supercooled Water: Molecular Origin of Dynamic and Thermodynamic Anomalies. *Physical Chemistry Chemical Physics* **2011**, *13*, 16220-16226.
84. Kim, J.; Torquato, S., Effect of Imperfections on the Hyperuniformity of Many-Body Systems. *Phys. Rev. B* **2018**, *97*, 054105.
85. Mahmud, G.; Zhang, H.; Douglas, J. F., Localization Model Description of the Interfacial Dynamics of Crystalline Cu and Cu64zr36 Metallic Glass Films. *Journal of Chemical Physics* **2020**, *153*, 124508.
86. Mahmud, G.; Zhang, H.; Douglas, J. F., The Dynamics of Metal Nanoparticles on a Supporting Interacting Substrate. *Journal of Chemical Physics* **2022**, *157*, 114505.
87. Plato; Gallop, D., *Phaedo*; Oxford University Press: Oxford ; New York, 1999, p xxix, 105 p.
88. Douglas, J. F.; Dudowicz, J.; Freed, K. F., Does Equilibrium Polymerization Describe the Dynamic Heterogeneity of Glass-Forming Liquids? *Journal of Chemical Physics* **2006**, *125*, 144907.
89. Dudowicz, J.; Freed, K. F.; Douglas, J. F., Lattice Model of Living Polymerization. Iii. Evidence for Particle Clustering from Phase Separation Properties and "Rounding" of the Dynamical Clustering Transition. *Journal of Chemical Physics* **2000**, *113*, 434-446.
90. Greer, S. C., Physical Chemistry of Equilibrium Polymerization. *J Phys Chem B* **1998**, *102*, 5413-5422.

91. Douglas, J. F.; Dudowicz, J.; Freed, K. F., Lattice Model of Equilibrium Polymerization. Vii. Understanding the Role of "Cooperativity" in Self-Assembly. *Journal of Chemical Physics* **2008**, *128*, 224901.
92. Dudowicz, J.; Freed, K. F.; Douglas, J. F., Lattice Model of Living Polymerization. I. Basic Thermodynamic Properties. *Journal of Chemical Physics* **1999**, *111*, 7116-7130.
93. Dudowicz, J.; Freed, K. F.; Douglas, J. F., Lattice Model of Equilibrium Polymerization. Iv. Influence of Activation, Chemical Initiation, Chain Scission and Fusion, and Chain Stiffness on Polymerization and Phase Separation. *Journal of Chemical Physics* **2003**, *119*, 12645-12666.
94. Gallo, P., et al., Water: A Tale of Two Liquids. *Chem Rev* **2016**, *116*, 7463-7500.
95. Neophytou, A.; Chakrabarti, D.; Sciortino, F., Topological Nature of the Liquid-Liquid Phase Transition in Tetrahedral Liquids. *Nature Physics, in press* **2022**, *18*, 1248-1253.
96. Holten, V.; Anisimov, M. A., Entropy-Driven Liquid-Liquid Separation in Supercooled Water. *Sci Rep-Uk* **2012**, *2*, 713.
97. Hocker, H.; Blake, G. J.; Flory, P. J., Equation-of-State Parameters for Polystyrene. *Transactions of the Faraday Society* **1971**, *67*, 2251.
98. Gong, L. L.; Zhang, X. Y.; Shi, Y., Investigation of Liquid-Liquid Transition Process in Atactic Polystyrene by Means of Thermally Stimulated Depolarization Current. *Eur. Polym. J.* **2011**, *47*, 1931-1935.
99. Lobanov, A. M.; Frenkel, S. Y., On the Nature of the So-Called "Liquid-Liquid" Transition in Polymer Melts. *Vysokomolekulyarnye Soedineniya Seriya A* **1980**, *22*, 1045-1057.
100. Maxwell, B.; Cook, K. S., Some Evidence for the Existence of the Liquid‐Liquid Transition: Melt Rheology Studies. *Journal of Polymer Science-Polymer Symposia* **1985**, 343-350.
101. Murthy, S. S. N., Liquid‐Liquid Transition in Polymers and Glass‐Forming Liquids. *J Polym Sci Pol Phys* **1993**, *31*, 475-480.
102. Wu, X. B.; Shang, S. Y.; Xu, Q. L.; Zhu, Z. G., Effects of Molecular Weight on the Liquid-Liquid Transition in Polystyrene Melts Studied by Low-Frequency Anelastic Spectroscopy. *J. Appl. Phys.* **2008**, *103*, 073519.
103. Shang, S. Y.; Zhu, Z. G.; Lu, Z. J.; Zhang, G. Z., Liquid-to-Liquid Relaxation of Polystyrene Melts Investigated by Low-Frequency Anelastic Spectroscopy. *Journal of Physics-Condensed Matter* **2007**, *19*, 416107.
104. Qian, R. Y.; Yu, Y. S., Transition of Polymers from Rubbery Elastic State to Fluid State. *Front. Chem. China* **2009**, *4*, 1-9.
105. Boyer, R. F., Contributions of Torsional Braid Analysis to Tu. *Polymer Engineering and Science* **1979**, *19*, 732-748.
106. Hedvat, S., Molecular Interpretation of T$_{//}$, the Rubbery-Viscous 'Transition' Temperature of Amorphous Polymers. *Polymer* **1981**, *22*, 774-777.
107. Wu, J. R.; Huang, G. S.; Pan, Q. Y.; Qu, L. L.; Zhu, Y. C.; Wang, B., Study on Liquid-Liquid Transition of Chlorinated Butyl Rubber by Positron Annihilation Lifetime Spectroscopy. *Appl Phys Lett* **2006**, *89*, 121904.
108. Kisliuk, A.; Mathers, R. T.; Sokolov, A. P., Crossover in Dynamics of Polymeric Liquids: Back to T-Parallel To? *J Polym Sci Pol Phys* **2000**, *38*, 2785-2790.
109. Boyer, R. F., Dynamics and Thermodynamics of the Liquid-State (T Greater-Than Tg) of Amorphous Polymers. *J Macromol Sci Phys* **1980**, *B18*, 461-553.
110. Fox, T. G.; Flory, P. J., 2nd-Order Transition Temperatures and Related Properties of Polystyrene .1. Influence of Molecular Weight. *J. Appl. Phys.* **1950**, *21*, 581-591.
111. Tanaka, F.; Matsuyama, A., Tricriticality in Thermoreversible Gels. *Phys. Rev. Lett.* **1989**, *62*, 2759-2762.
112. Douglas, J. F.; Hubbard, J. B., Semiempirical Theory of Relaxation - Concentrated Polymer-Solution Dynamics. *Macromolecules* **1991**, *24*, 3163-3177.

113. Chremos, A.; Douglas, J. F., Communication: When Does a Branched Polymer Become a Particle? *Journal of Chemical Physics* **2015**, *143*, 111104.
114. Chremos, A.; Douglas, J. F., A Comparative Study of Thermodynamic, Conformational, and Structural Properties of Bottlebrush with Star and Ring Polymer Melts. *Journal of Chemical Physics* **2018**, *149*, 044904.
115. Vargas-Lara, F.; Betancourt, B. A. P.; Douglas, J. F., Communication: A Comparison between the Solution Properties of Knotted Ring and Star Polymers. *Journal of Chemical Physics* **2018**, *149*, 161101.
116. Vargas-Lara, F.; Hassan, A. M.; Mansfield, M. L.; Douglas, J. F., Knot Energy, Complexity, and Mobility of Knotted Polymers. *Sci Rep-Uk* **2017**, *7*, 13374.
117. Vargas-Lara, F.; Betancourt, B. A. P.; Douglas, J. F., Influence of Knot Complexity on Glass-Formation in Low Molecular Mass Ring Polymer Melts. *Journal of Chemical Physics* **2019**, *150*, 101103.
118. Chremos, A.; Horkay, F.; Douglas, J. F., Influence of Network Defects on the Conformational Structure of Nanogel Particles: From "Closed Compact" to "Open Fractal" Nanogel Particles. *Journal of Chemical Physics* **2022**, *156*, 094903.
119. Chremos, A.; Horkay, F.; Douglas, J. F., Structure and Conformational Properties of Ideal Nanogel Particles in Athermal Solutions. *Journal of Chemical Physics* **2021**, *155*, 134905.
120. Betancourt, B. A. P.; Douglas, J. F.; Starr, F. W., String Model for the Dynamics of Glass-Forming Liquids. *Journal of Chemical Physics* **2014**, *140*, 204509.
121. Betancourt, B. A. P.; Starr, F. W.; Douglas, J. F., String-Like Collective Motion in the Alpha- and Beta-Relaxation of a Coarse-Grained Polymer Melt. *Journal of Chemical Physics* **2018**, *148*, 104508.
122. Hopkins, A. B.; Stillinger, F. H.; Torquato, S., Nonequilibrium Static Diverging Length Scales on Approaching a Prototypical Model Glassy State. *Phys. Rev. E* **2012**, *86*, 021505.
123. Karmakar, S.; Dasguptaa, C.; Sastry, S., Growing Length and Time Scales in Glass-Forming Liquids. *P Natl Acad Sci USA* **2009**, *106*, 3675-3679.
124. Wei, S.; Coleman, G. J.; Lucas, P.; Angell, C. A., Glass Transitions, Semiconductor-Metal Transitions, and Fragilities in Ge-V-Te (V = as, Sb) Liquid Alloys: The Difference One Element Can Make. *Phys. Rev. Appl.* **2017**, *7*, 034035.
125. Du, Q., et al., Reentrant Glass Transition Leading to Ultrastable Metallic Glass. *Mater Today* **2020**, *34*, 66-77.
126. Shen, J., et al., Metallic Glacial Glass Formation by a First-Order Liquid-Liquid Transition. *J. Phys. Chem. Lett.* **2020**, *11*, 6718-6723.
127. Orava, J.; Greer, A. L.; Gholipour, B.; Hewak, D. W.; Smith, C. E., Characterization of Supercooled Liquid Ge2sb2te5 and Its Crystallization by Ultrafast-Heating Calorimetry. *Nat Mater* **2012**, *11*, 279-283.
128. Orava, J.; Weber, H.; Kaban, I.; Greer, A. L., Viscosity of Liquid Ag-in-Sb-Te: Evidence of a Fragile-to-Strong Crossover. *Journal of Chemical Physics* **2016**, *144*, 194503.
129. Steimer, C.; Coulet, V.; Welnic, W.; Dieker, H.; Detemple, R.; Bichara, C.; Beuneu, B.; Gaspard, J. P.; Wuttig, M., Characteristic Ordering in Liquid Phase-Change Materials. *Advanced Materials* **2008**, *20*, 4535-4540.

# Supplementary Information:
# Approach to Hyperuniformity in a Metallic Glass-Forming Material Exhibiting a Fragile to Strong Glass Transition

*Hao Zhang[1†], Xinyi Wang[1], Jiarui Zhang[1], Hai-Bin Yu[2], Jack F. Douglas[3†]*

[1] Department of Chemical and Materials Engineering, University of Alberta, Edmonton, Alberta, Canada, T6G 1H9

[2] Wuhan National High Magnetic Field Center, Huazhong University of Science and Technology, Wuhan, Hubei, China, 430074

[3] Material Measurement Laboratory, Material Science and Engineering Division, National Institute of Standards and Technology, Maryland, USA, 20899

*Corresponding authors: hao.zhang@ualberta.ca; jack.douglas@nist.gov

**A. Fragile-Strong Transition, Liquid-liquid Transition, and Equilibrium Polymerization**

As we discuss in some detail below, many cooled GF fluids seem to exhibit the formation of linear polymeric clusters whose size grows upon cooling. This situation makes the theory and simulations of Sciortino and coworkers [1] of the molecular clustering of fluid particles forming dynamic linear polymer chains of obvious relevance to modeling the low-$q$ upturn in $S(q)$ of GF liquids and other complex fluids in which this type of dynamic clustering is prevalent. As anticipated, a direct comparison of $S(q)$ data for a simulated fluid undergoing equilibrium polymerization (see Fig. 13 of Sciortino et al.) with light scattering measurements of $S(q)$ on orthoterphenyl, a model GF liquid of the OGF type [2], upon approaching its $T_g$, reveals a remarkable resemblance, and in many measurements on other GF systems, this type of scattering reveals itself as a low-$q$ upturn with no obvious tendency of the scattering intensity to saturate to finite value at very low $q$ [2-3], presumably due to the large size of the polymeric clusters forming in these systems in comparison to the wavelength of the scattering radiation probing them. This "anomalous" scattering in GF liquids has traditionally been attributed to mysterious "Fisher clusters" [2-3], named in honor of the scientist who most extensively studied this striking phenomenon. Despite its common occurrence, this conspicuous scattering feature of GF liquids is often ignored because of any traditional interpretation of this phenomenon. We suggest that this type of scattering data provides direct evidence about the formation of polymeric clusters in cooled liquids, which notably do not necessarily exist in a state of equilibrium in real cooled liquids because of the extremely long times required for equilibration to occur in such self-assembly processes. It is stressed that this type of structure formation and the corresponding $S(q)$ obtained from such clusters do not give rise to a pre-peak feature in $S(q)$. This scattering feature, which is found in some but not all GF liquids, is telling us that the structural organization in the liquid is more complex than chain-like structures having a linear topology.

The formation of dynamic polymer chains is just one "universality class" of equilibrium polymerization. More generally, associating molecular and atomic species (specific examples considered below) capable of multi-valent associations normally results in the formation of randomly branched equilibrium polymers in which the polymer size distribution and fractal geometry of the polymeric clusters are *completely different* from the case of linear chain formation at equilibrium. [4-5] This is a physical situation with network-forming GF liquids, such as water and silica, and this type of dynamic self-assembly process occurs in numerous supramolecular processes in aqueous solutions that can potentially inform their counterpart in GF liquids. Finally, for completeness, we mention the third general class of equilibrium self-assembly processes in which the assemblies take the form of compact objects such as spherical micelles and spherical viral capsid structures.

Unfortunately, there is no exact analytic theory of $S(q)$ for fluids exhibiting this common type of supramolecular assembly, but it is possible to investigate this type of self-assembly process by molecular dynamics, Brownian dynamics or Monte Carlo simulation. The scientific literature on $S(q)$ in this type of self-assembling system is extensive, but a full understanding of how to model $S(q)$ does not currently exist. We can nonetheless understand some of the general trends in the scattering data shown in Fig. 1 based on this type of simulation study and from scattering measurements on model randomly branched polymer materials. Most basically, $S(q)$ in this type of branched polymer structure exhibits a low-$q$ upturn as found for linear polymers, but the scattering from these structures often exhibits a pre-peak feature at higher $q$ reflecting the mesh structure of the network or other characteristic dimension of the branched polymer. This type of structural organization is common in simulations and measurements of $S(q)$ made on polyelectrolyte solutions [6-9] and bottlebrush polymers. [10] The importance of this class of polymer solutions has led to simulations of $S(q)$ of model network polymers [8-9, 11-12] to gain insight into these scattering features in these complex fluids and how these scattering features relate to the molecular geometry of branched polymers.

The pre-peak generally corresponds to some type of mesoscale organization of the polymeric network, but it has been difficult to tie down a unique interpretation that applies to all branched polymeric structures. In the future, we plan to further simulate model networks of different types, formed by self-assembly, or as structures with an assumed static topological structure, to better understand the structural origin of the pre-peak in diverse materials in which randomly branched polymeric structures arise. We next more discuss the nature of equilibrium polymerization in particular GF liquids in a chemically specific way.

There has been intense research recently aimed at quantifying structural organization in metallic GF materials [13-17] and computational and ultrahigh-resolution measurement studies have consistently indicated that the "structuring" in these materials involves a kind of short-range ordering (SRO) at atomic scales in which larger "solute" atoms, such as our Sm atoms in our Al-Sm metallic glass system, are 'solvated' by the smaller atomic species to form well-defined clusters in the metallic GF liquid having an approximate local icosahedral symmetry, while the structure formation at larger scales corresponding to "medium range ordering" (MRO) in the material corresponds to *polymeric structures* comprised of the primary icosahedral clusters. The physical situation in metallic GF liquids at low $T$ then resembles the hierarchical assembly of worm-like micelles and amyloid fibers in aqueous solutions in which compact clusters first form, and then these structures self-assemble in turn to form *polymeric assembles* that coexist with the smaller clusters. [18-23] In particular, both measurements and simulations have revealed in the *strings of icosahedra* form and these structures organize into domains at a larger scale typically on the order of a few nm. These relatively "ordered" regions are surrounded by relatively "loosely-packed" regions [14, 16-17] that are largely "disordered" and lower in density, in addition to cavity regions devoid of particles altogether. Nanoprobe measurements of the local elasticity of metallic glass materials have provided additional insights into this nanoscale heterogeneity. [24-25] These observations are broadly consistent with the occurrence of a kind of microphase separation in which there are coexisting phases of distinct entropy, a phenomenon investigated in many works on metallic and metallic oxide GF materials, [26-28] organic GF liquids [29-31] and discussed intensively recently in many simulations and experimental studies in connection with understanding the thermodynamic and dynamic properties of water at low $T$. [32-35] We examined this type of structure formation in our earlier work on the Al-Sm metallic glass system. [36-37]

**B. Specific Examples of Fluids Exhibiting Equilibrium Polymerization**

We note that there is evidence of "string" formation of this kind even in single-component atomic fluids, which likewise exhibit FS glass formation and microphase phase separation. In particular, this type of *equilibrium polymerization* process arises in liquids of chalcogenide elements where there is a propensity for two-fold coordination because of the two "lone pair" electrons of these elements. [38] Equilibrium polymerization has also been suggested in other elements such as C under conditions where $sp^3$ hybridization of the atomic orbitals is prevalent.[39] The literature is very extensive, and here we simply mention some representative references: S [40-43], P [38, 44], and Se. [45-47] Simulation studies indicate that this phenomenon arises in C [39], Te [48-51], Ge [52], Ga [53-54] and other elements and recent measurements [55] on supercooled liquid Te have indicated a pre-peak in the structure factor, along with anomalies in its thermodynamic response functions that are similar to those observed in water and our Al-Sm glass-forming liquid.

Although the propensity of the atomic species to form polymeric structures upon cooling is inherently linked to the chemistry of the atomic species in atomic fluids, the formation of polymeric icosahedral clusters is also commonly observed in simulations [56] of hard sphere fluids. Such clusters have apparently been observed in model granular "hard sphere" fluids. [57] Many-body effects can evidently also give rise to this type of self-assembly process due to a purely entropic driving force. Remarkably, this tendency towards anisotropic bonding arises even though the individual particles do not possess any anisotropy in their intermolecular potential.

We note that the tendency of molecules and particles to form dynamic polymeric structures at low $T$ can be modelled roughly by a simple two-state model corresponding to an associated or "polymeric" species and the un-associated or "monomer" species as a coarse-grained model that neglects the polydispersity in this type of dynamic heterogeneity. [58] There is a long history of invoking this type of simplified model to describe the dynamics of water and other network forming liquids, and Tanaka and coworkers [59-62] have developed an interesting variant of this type of model based on the assumption of the dynamic coexistence between water molecules participating in locally energetically preferred structures and water molecules existing in a "normal" liquid-like state. Despite the neglect of the specific structural form of the locally preferred structures, their remarkably simple model is apparently able to capture many aspects of the thermodynamics and dynamics of water, including the "anomalies" in thermodynamic response functions of the type shown in Fig. 3 of our paper and the striking change from a high-$T$ Arrhenius regimes to a low-$T$ Arrhenius regime of diffusion and structural relaxation that is apparently characteristic of liquids exhibiting fragile-to strong glass-formation . Douglas et al. [58] have shown that the variable cooperativity of equilibrium polymerization transition that arise from *constraining* the chain of particle association events in the polymer formation process can be quantitatively emulated by varying the order *m* of the associative reaction process in the mean two-state model of thus type dynamic structure formation, this type of model having long-standing applications as a simplified model for micelle formation and formation of various biological assemblies.[58] Variable cooperativity in a similar mathematical sense is also basic feature of glass-formation,[63] and we discuss this basic property of thermodynamic polymerization transitions below since it strongly relates to the propensity to whether a fluid exhibits liquid-liquid phase separation. While the simple two-state model is highly convenient for understanding many qualitative aspects of fluids exhibiting dynamic polymeric assembly there are certain phenomena in which the specific geometrical form and structural polydispersity of the "locally energetically preferred structures" discussed by Tanaka and coworkers [59, 62] (dynamic polymers in the language of the present work) can be expected to become of significant importance. We next discuss how the propensity for liquid-liquid phase separation can be understood from an equilibrium polymerization perspective as an example of this structural sensitivity.

The analytic theory of fluids undergoing equilibrium polymerization, in conjunction with phase separation, reveals an interesting (arguably "strange') many-body phenomenon that can arise from the coupling of these distinct thermodynamic transitions which seems to be directly relevant to systems exhibiting FS glass-formation. In particular, when the polymerization transition is highly cooperative in the precise sense of approaching a second-order phase transition due to thermal activation or initiation by a chemical species,[58, 64] the phase boundary can develop *two* critical points, one critical point in the vicinity of the transition near the fluid without the associative interaction giving rise to equilibrium polymerization and a second "liquid-liquid" critical point at much lower temperatures and at a greatly altered critical concentration corresponding to where the polymerization transition line intersects the phase boundary of the "primary" critical point. We will discuss the polymerization transition line and its possible significance for FS GF liquids below. Interestingly, the lower liquid-liquid critical point involves the coexistence of polymers of different topologies. The equilibrium polymer theory just mentioned refers to the restricted case of the formation of linear equilibrium polymers, but multiple critical points also arise by an apparently similar mechanism in fluids in which randomly branched polymers form by equilibrium polymerization. [65] This is potentially relevant to the current discussion because Naserifar and Goddard [66] have reported based on first-principle quantum calculations that liquid water, a model fluid exhibiting FS glass formation, can be described as "polydisperse branched polymer" material. Recent simulation observations by Sciortino and coworkers [67] have indicated that the high and low density forms of water associated with low-$T$ liquid-liquid phase separation in this fluid corresponds to the hierarchical assembly into tetrahedral water clusters, which in turn self-assemble at a larger scale into polydisperse branched polymers having distinct topological complexity, which is highly reminiscent of observations of hierarchical polymerization of "superstructures" in metallic glasses, mentioned above. We hypothesize based on these theoretical findings and simulation observations that the high fragility of glass-formers at elevated temperatures characteristic of glass-formers exhibiting FS glass-formation should naturally engender a tendency toward liquid-liquid phase transition between self-assembled structures of different topology at lower temperatures as a general attribute of GF liquids exhibiting FS glass-formation. This situation also allows understanding in a unified way why GF liquids exhibiting less cooperativity should not exhibit liquid-liquid phase separation. Betancourt et al.

have suggested a possible physical mechanism modulating the cooperativity of the polymerization transition in GF liquids. [68]

Detailed calculations indicate that the occurrence of *multiple* critical points in systems undergoing *linear chain polymerization* only occurs under a rather special set of conditions in which the polymerization transition is *highly cooperative* and the magnitude of the short-range van der Waals interaction to the interaction strength of the interaction giving rise to the formation of polymeric structures is relatively weak. [64] Water, for example, would seem to satisfy these conditions since this fluid has a relatively high cohesive energy density, and water molecules have relatively large dipolar and quadrupolar interactions that naturally engender a tendency toward directional self-assembly into polymeric structures. Cooperativity in equilibrium polymerization[58] can be quantified similarly to GF liquids in terms of the rate of change of the configurational entropy when $T$ is varied, where this variation in thermodynamics bears a direct relationship to the formation of polymers as a kind of structural "ordering". [63] [68-71]

The impact of this type of equilibrium polymerization process on the viscoelastic and stress relation of GF liquids has been emphasized by Douglas and Hubbard, [72] but this theory for relaxation and creep under steady stress remains to be quantitatively tested. This theory predicts many observed aspects of GF liquids such as $\alpha$-$\beta$ bifurcation, decoupling, stretched exponential relaxation, Andrade creep, etc. [72-74] Previous measurements have demonstrated the predictive power of standard polymer models for quantitatively understanding the viscoelasticity of cooled liquid metallic materials with a known propensity for equilibrium polymerization (e.g., Se, Te) [75-76], and there would appear to considerable scope for this approach given the ubiquity of the observation of polymerization processes in association with the ordering of metallic glass and other GF materials.

Interestingly, the idea that polymerization underlies glass-formation can be traced back to the advent of the theoretical study of GF liquids. In particular, Hägg [77] in 1934 suggested this hypothesis in response to Zachariasen's introduction of the "random network model" of glasses.[78] He suggested that rather than the atomic positions being truly random, the atoms of both metallic and inorganic glasses form one-dimensional collineations (or two-dimensional sheet polymers or highly perforated sheets, i.e., branched polymers[12, 79]), often composed tetrahedra in ion complexes or arising from the bonding habits of low coordination in "metalloid" elements such as the chalcogenides. In modern terminology, Hägg suggested that the purely random configuration of particles is energetically unstable to adopt local energetically preferred structures in the liquid that cannot be shared by all the particles because of packing frustration so that the material adopts a correlated heterogeneous structure in which string and sheet-like structures arise. The same driving force of energy minimization drives the formation of crystalline materials, but the energy minimization applies to all particles so that the energy of the system is invariant to permuting the particle positions, a basic attribute of crystalline materials. The exchange symmetry is only approximately true for quasi-crystals except for a set of exceptional atomic defect suites that allow for ordered structures having symmetries (e.g., five-fold rotational symmetry) consistent with perfectly periodic crystals. The same situation holds for the formation sheaths of proteins in viruses where the ordered structures lacking a perfect exchange symmetry, as suggested by Crick and Watson, [80] are replaced by "quasi-equivalent" structures [81-84] in which the defects are distributed to minimize the energy of the "crystalline" structure to form structures having otherwise disallowed symmetries. In a sense, quasi-crystals can be viewed as "quasi-equivalent crystals". Hägg also argued that these structures first formed within the liquid state and grew progressively in size upon cooling, thereby frustrating crystallization because of the reduction of particle mobility of molecular species localized in the clusters and by topological interactions caused by the "muddle of other chains" that further inhibited atomic movement, i.e., entanglement in modern terms. Apart from the prescience of this model of glass-formation, Hägg's model is especially remarkable given that the concept of polymers was hardly accepted at that time. After a heated exchange between Zachariasen and Hägg, [85-86] his model, unfortunately, sank into oblivion until recently when measurements have largely confirmed his conception of a specific form of "dynamic heterogeneity" that is found in many GF liquids (Evidence is discussed below.).

**C. Equilibrium Polymerization on Glass-Formation in Multiple Component Materials**

This polymerization model of glass-formation, in which the "ordering" process in many GF liquids is viewed as a form of equilibrium polymerization [63, 72], also has strong implications for the miscibility of metal alloys and other multi-component GF liquids, since polymerization inherently alters the miscibility of mixtures and can lead to a multiplicity of critical points arising from competition of the polymerization thermodynamics and phase separation. [64, 87-88] This issue provides an added level of complexity to metallic GF alloys since the liquid-liquid phase separation between liquid structures having different topologies, as discussed above, should occur simultaneously with ordinary liquid-liquid phase separation between different chemical species. A change in miscibility arising from equilibrium polymerization can be particularly significant when one of the components self-assembles upon heating, as in the case of S. This type of additive is predicted [89] to give rise to closed loop phase behavior, as observed in S-Te mixtures. [90-93]

Tsuchiya [87] explored the relevance of this type of "ordering process" on the thermodynamic stability of Te alloys and he also mentions[87] this phenomenon in S solutions [41] (See Fig. 1a of Ref. 41) and in $He^3$-$He^4$ mixtures [94-95]. Note that the liquid-liquid transition between the normal and superfluid liquid states occurs as a line of second-order phase transitions emanating from the critical point in the He mixtures (a tricritical point). This transition can also be viewed as a polymerization transition line based on Feynman's equilibrium polymerization model of the superfluid transition where the polymers were assumed to involve the collective motion of atoms moving permutationally in the form of ring polymers. [96-97] In our previous work on the Al-Sm metallic glass [36-37], we observed a similar permutational motion in our classical molecular dynamics of the Al-Sm GF liquid, although these dynamic polymer chains (which we call "stings") were not found to be generally closed to form rings. [36] Moreover, these dynamic polymeric structures were shown to be relevant to understanding the $T$ dependence of mass diffusion and the relaxation times of this metallic GF liquid.[98]

The phase behavior of associative fluids often involves a line of polymerization or self-assembly transitions that terminate near the phase boundary for liquid-liquid phase separation, giving rise to a tricritical point when the line of transitions intersects the phase boundary at the critical point and a critical endpoint when the polymerization line intersects the binodal of the phase boundary away from the critical point. Importantly, this phenomenon can arise both when the polymerization process involves linear chains [64, 99] and when the polymerization process involves randomly branched polymeric structures. [65] This phenomenon is ubiquitous in complex fluids in which molecular assembly occurs, including micelle formation [100-101], living polymers [41], thermally reversible gelation of polymers in solution [102-103], etc. Liquid-liquid thermodynamic transitions, accompanied by significant changes in fluid structure and viscoelastic properties in the two liquid states, are common in complex fluids, even if this phenomenon is not always recognized. Tanaka has recently provided an up-to-date review of the exotic phenomenon of liquid-liquid phase separation in single component fluids.[104]

**D. Suggested Physical Origin of the Widom Line**

One of the important features of liquid-liquid phase transitions in fluids exhibiting equilibrium polymerization is that the transition from a monomeric to a viscous polymeric liquid may occur as a rounded second-order thermodynamic transition upon crossing the polymerization line in the one-phase region can be first-order under conditions in which the polymerization line intersects the phase boundary [58, 105] (See Fig. 1 of Ref. [99]). The polymerization line often extends beyond the phase boundary, forming an "extension line" into the one-phase region at which the correlation length for composition fluctuations peaks, reflecting the clustering of fluid molecules near the polymerization line. [58, 99] In particular, the polymeric particle clusters become large upon approaching the polymerization line and the scattering function takes the form appropriate for Ornstein-Zernike form for flexible polymers where the correlation length average polymer size grows in parallel. [105]

The "Widom line" [106-107] is often identified in the liquid state literature with a locus in the temperature-density plane at which the correlation length and scattering intensity of a liquid exhibit maximum in the one-phase region, forming a line in this parameter space that typically intersects the phase boundary in the vicinity of the critical point, at which these basic scattering properties diverge. In practice, however, the Widom line is identified in studies of the water, and other liquids thought to exhibit liquid-liquid phase separation, [106, 108] by a characteristic temperature at which $C_p$ exhibits a maximum where a line of such characteristic $T$ is obtained by varying the density through changes in the applied pressure. Liquid-liquid transitions are defined by a peak in $C_p$ in fluids exhibiting supramolecular polymerization. This phenomenon is well exemplified in the Stockmayer fluid [109-110], a simplified model of water that includes a Lennard-Jones interaction in combination with a superimposed dipolar interaction that gives rise to chain assembly. The appearance of a "ridge" in the supercritical region has been observed since the 1930s when some type of transition between the gas-like and liquid-like organization and molecular clustering was inferred by x-ray scattering, sound propagation and other measurements and this type of observation was organized and emphasized by Nishikawa and coworkers. [111-114] Simulations have recently indicated that this line of transitions does not exist in mixtures of binary Lennard-Jones fluids [115] so the directional interaction is required for self-assembly and the observation of the transition line. We suggest that the extension line of supercooled liquids may be identified with a polymerization line so that the termination of this transition is appropriately made from a maximum in $C_p$, the condition utilized by Stanley and coworkers for experimentally locating the Widom line. The existence of a polymerization transition in cooled liquids offers a potentially new interpretation of the Widom line.

**E. Relation Between FS Glass-Formers and Superionic Crystalline Materials?**

Angell [116-117] has suggested that GF liquids undergoing an FS-type transition, including water, silica, $BeF_2$, and some metallic GF materials, are in some ways "crystal-like in character while remaining strictly aperiodic", and he further suggested that these materials are somehow intermediate between quasi-crystals and ordinary GF liquids. We interpret these impressionistic thoughts as being consistent with effectively hyperuniform materials which are "crystal-like" specifically because of their relatively low values of $H$. [118-119] Based on this intuitive conception of these materials, he introduced a "bond lattice" model [117, 120-121] to rationalize both thermodynamic and relaxation properties of these odd GF liquids having little or no evidence of a peak in the specific heat as the material goes out of equilibrium, a defining feature of the glass-transition in many experimental studies of conventional materials exhibiting glass-formation, while these materials are also peculiar in that they exhibit a thermodynamic "lambda transition", characterized by a peak in the specific heat resembling a thermodynamic melting transition, as found in crystalline materials. This is another way these materials are "crystal-like" that his model addresses. He went on to speculate that "plastic crystals" exhibiting orientational disorder and superionic crystalline materials and globular proteins might belong to this same general class of materials. In accord with these remarkable suggestions, we have observed striking similarities between the dynamics of our Al-Sm metallic glass and simulations of the dynamics of superionic $UO_2$. In our previous simulations of $UO_2$, we noticed that the $T$-dependent dynamics of this material has a strong resemblance to a singular $T$-dependence of the rate of certain enzyme reactions.[122] Angell's ideas regarding the FS thus seem to have a lot of merits, of course, he does not offer any clear underlying physical reason for this distinct type of glass-formation that might allow some understanding of why all these materials are "related".

# References

1. Sciortino, F.; Bianchi, E.; Douglas, J. F.; Tartaglia, P., Self-Assembly of Patchy Particles into Polymer Chains: A Parameter-Free Comparison between Wertheim Theory and Monte Carlo Simulation. *Journal of Chemical Physics* **2007**, *126*, 194903.
2. Patkowski, A.; Thurn-Albrecht, T.; Banachowicz, E.; Steffen, W.; Bosecke, P.; Narayanan, T.; Fischer, E. W., Long-Range Density Fluctuations in Orthoterphenyl as Studied by Means of Ultrasmall-Angle X-Ray Scattering. *Phys. Rev. E* **2000**, *61*, 6909-6913.
3. Fischer, E. W., Light-Scattering and Dielectric Studies on Glass-Forming Liquids. *Physica A* **1993**, *201*, 183-206.
4. Audus, D. J.; Starr, F. W.; Douglas, J. F., Coupling of Isotropic and Directional Interactions and Its Effect on Phase Separation and Self-Assembly. *Journal of Chemical Physics* **2016**, *144*, 074901.
5. Audus, D. J.; Starr, F. W.; Douglas, J. F., Valence, Loop Formation and Universality in Self-Assembling Patchy Particles. *Soft Matter* **2018**, *14*, 1622-1630.
6. Chremos, A.; Douglas, J. F., Communication: Counter-Ion Solvation and Anomalous Low-Angle Scattering in Salt-Free Polyelectrolyte Solutions. *Journal of Chemical Physics* **2017**, *147*, 241103.
7. Chremos, A.; Douglas, J. F., Polyelectrolyte Association and Solvation. *Journal of Chemical Physics* **2018**, *149*, 163305.
8. Horkay, F.; Chremos, A.; Douglas, J. F.; Jones, R. L.; Lou, J. Z.; Xia, Y., Systematic Investigation of Synthetic Polyelectrolyte Bottlebrush Solutions by Neutron and Dynamic Light Scattering, Osmometry, and Molecular Dynamics Simulation. *Journal of Chemical Physics* **2020**, *152*, 194904.
9. Horkay, F.; Chremos, A.; Douglas, J. F.; Jones, R.; Lou, J. Z.; Xia, Y., Comparative Experimental and Computational Study of Synthetic and Natural Bottlebrush Polyelectrolyte Solutions. *Journal of Chemical Physics* **2021**, *155*, 074901.
10. Sarapas, J. M.; Martin, T. B.; Chremos, A.; Douglas, J. F.; Beers, K. L., Bottlebrush Polymers in the Melt and Polyelectrolytes in Solution Share Common Structural Features. *P Natl Acad Sci USA* **2020**, *117*, 5168-5175.
11. Chremos, A.; Horkay, F.; Douglas, J. F., Structure and Conformational Properties of Ideal Nanogel Particles in Athermal Solutions. *Journal of Chemical Physics* **2021**, *155*, 134905.
12. Chremos, A.; Douglas, J. F.; Basser, P. J.; Horkay, F., Molecular Dynamics Study of the Swelling and Osmotic Properties of Compact Nanogel Particles. *Soft Matter* **2022**, *18*, 6278-6290.
13. Sheng, H. W.; Luo, W. K.; Alamgir, F. M.; Bai, J. M.; Ma, E., Atomic Packing and Short-to-Medium-Range Order in Metallic Glasses. *Nature* **2006**, *439*, 419-425.
14. Li, M.; Wang, C. Z.; Hao, S. G.; Kramer, M. J.; Ho, K. M., Structural Heterogeneity and Medium-Range Order in Zrxcu100-X Metallic Glasses. *Phys. Rev. B* **2009**, *80*, 184201.
15. Yavari, A. R., Materials Science - a New Order for Metallic Glasses. *Nature* **2006**, *439*, 405-406.
16. Hirata, A.; Guan, P. F.; Fujita, T.; Hirotsu, Y.; Inoue, A.; Yavari, A. R.; Sakurai, T.; Chen, M. W., Direct Observation of Local Atomic Order in a Metallic Glass. *Nat Mater* **2011**, *10*, 28-33.
17. Zhu, F.; Hirata, A.; Liu, P.; Song, S. X.; Tian, Y.; Han, J. H.; Fujita, T.; Chen, M. W., Correlation between Local Structure Order and Spatial Heterogeneity in a Metallic Glass. *Phys. Rev. Lett.* **2017**, *119*, 215501.
18. Jensen, G. V.; Lund, R.; Gummel, J.; Narayanan, T.; Pedersen, J. S., Monitoring the Transition from Spherical to Polymer-Like Surfactant Micelles Using Small-Angle X-Ray Scattering. *Angew. Chem.-Int. Edit.* **2014**, *53*, 11524-11528.
19. Sun, P. P.; Lu, F.; Wu, A. L.; Shi, L. J.; Zheng, L. Q., Spontaneous Wormlike Micelles Formed in a Single-Tailed Zwitterionic Surface-Active Ionic Liquid Aqueous Solution. *Soft Matter* **2017**, *13*, 2543-2548.
20. Schurtenberger, P.; Cavaco, C.; Tiberg, F.; Regev, O., Enormous Concentration-Induced Growth of Polymer-Like Micelles. *Langmuir* **1996**, *12*, 2894-2899.

21. Jain, S.; Udgaonkar, J. B., Evidence for Stepwise Formation of Amyloid Fibrils by the Mouse Prion Protein. *J Mol Biol* **2008**, *382*, 1228-1241.
22. Dudowicz, J.; Freed, K. F.; Douglas, J. F., Lattice Model of Living Polymerization. Iii. Evidence for Particle Clustering from Phase Separation Properties and "Rounding" of the Dynamical Clustering Transition. *Journal of Chemical Physics* **2000**, *113*, 434-446.
23. Ross, C. A.; Poirier, M. A., Protein Aggregation and Neurodegenerative Disease. *Nat. Med.* **2004**, *10*, S10-S17.
24. Liu, Y. H.; Wang, D.; Nakajima, K.; Zhang, W.; Hirata, A.; Nishi, T.; Inoue, A.; Chen, M. W., Characterization of Nanoscale Mechanical Heterogeneity in a Metallic Glass by Dynamic Force Microscopy. *Phys. Rev. Lett.* **2011**, *106*, 125504.
25. Liu, C. Y.; Maass, R., Elastic Fluctuations and Structural Heterogeneities in Metallic Glasses. *Advanced Functional Materials* **2018**, *28*, 1800388.
26. Aasland, S.; McMillan, P. F., Density-Driven Liquid–Liquid Phase Separation in the System Ai2o3–Y2o3. *Nature* **1994**, *369*, 633-636.
27. Wei, S.; Yang, F.; Bednarcik, J.; Kaban, I.; Shuleshova, O.; Meyer, A.; Busch, R., Liquid-Liquid Transition in a Strong Bulk Metallic Glass-Forming Liquid. *Nat. Commun.* **2013**, *4*, 2083.
28. Shen, J., et al., Metallic Glacial Glass Formation by a First-Order Liquid-Liquid Transition. *J. Phys. Chem. Lett.* **2020**, *11*, 6718-6723.
29. Tanaka, H.; Kurita, R.; Mataki, H., Liquid-Liquid Transition in the Molecular Liquid Triphenyl Phosphite. *Phys. Rev. Lett.* **2004**, *92*, 025701.
30. Zhu, M.; Wang, J. Q.; Perepezko, J. H.; Yu, L., Possible Existence of Two Amorphous Phases of D-Mannitol Related by a First-Order Transition. *Journal of Chemical Physics* **2015**, *142*, 244504.
31. Kobayashi, M.; Tanaka, H., The Reversibility and First-Order Nature of Liquid-Liquid Transition in a Molecular Liquid. *Nat. Commun.* **2016**, *7*, 13438.
32. Poole, P. H.; Grande, T.; Angell, C. A.; McMillan, P. F., Polymorphic Phase Transitions in Liquids and Glasses. *Science* **1997**, *275*, 322-323.
33. Mishima, O.; Stanley, H. E., The Relationship between Liquid, Supercooled and Glassy Water. *Nature* **1998**, *396*, 329-335.
34. Debenedetti, P. G.; Sciortino, F.; Zerze, G. H., Second Critical Point in Two Realistic Models of Water. *Science* **2020**, *369*, 289.
35. Kringle, L.; Thornley, W. A.; Kay, B. D.; Kimmel, G. A., Reversible Structural Transformations in Supercooled Liquid Water from 135 to 245 K. *Science (New York, N.Y.)* **2020**, *369*, 1490-1492.
36. Zhang, H.; Wang, X. Y.; Yu, H. B.; Douglas, J. F., Dynamic Heterogeneity, Cooperative Motion, and Johari-Goldstein Beta-Relaxation in a Metallic Glass-Forming Material Exhibiting a Fragile-to-Strong Transition. *Eur Phys J E* **2021**, *44*.
37. Zhang, H.; Wang, X. Y.; Yu, H. B.; Douglas, J. F., Fast Dynamics in a Model Metallic Glass-Forming Material *Journal of Chemical Physics* **2021**, *154*, 084505.
38. Katayama, Y.; Mizutani, T.; Utsumi, W.; Shimomura, O.; Yamakata, M.; Funakoshi, K., A First-Order Liquid-Liquid Phase Transition in Phosphorus. *Nature* **2000**, *403*, 170-173.
39. Glosli, J. N.; Ree, F. H., Liquid-Liquid Phase Transformation in Carbon. *Phys. Rev. Lett.* **1999**, *82*, 4659-4662.
40. Tobolsky, A. V.; Eisenberg, A., Equilibrium Polymerization of Sulfur. *J Am Chem Soc* **1959**, *81*, 780-782.
41. Greer, S. C., Physical Chemistry of Equilibrium Polymerization. *J Phys Chem B* **1998**, *102*, 5413-5422.
42. Zhang, L. J.; Ren, Y.; Liu, X. R.; Han, F.; Evans-Lutterodt, K.; Wang, H. Y.; He, Y. L.; Wang, J. L.; Zhao, Y.; Yang, W. G., Chain Breakage in the Supercooled Liquid - Liquid Transition and Re-Entry of the Lambda-Transition in Sulfur. *Sci Rep-Uk* **2018**, *8*, 4558.
43. Henry, L.; Mezouar, M.; Garbarino, G.; Sifre, D.; Weck, G.; Datchi, F., Liquid-Liquid Transition and Critical Point in Sulfur. *Nature* **2020**, *584*, 382.

44. Hohl, D.; Jones, R. O., Polymerization in Liquid Phosphorus: Simulation of a Phase Transition. *Phys. Rev. B* **1994**, *50*, 17047-17053.
45. Bohmer, R.; Angell, C. A., Elastic and Viscoelastic Properties of Amorphous Selenium and Identification of the Phase Transition between Ring and Chain Structures. *Phys. Rev. B* **1993**, *48*, 5857-5864.
46. Bichara, C.; Pellegatti, A.; Gaspard, J. P., Chain Structure of Liquid Selenium Investigated by a Tight-Binding Monte Carlo Simulation. *Phys. Rev. B* **1994**, *49*, 6581-6586.
47. Kalikka, J.; Akola, J.; Jones, R. O.; Schober, H. R., Density Functional and Classical Simulations of Liquid and Glassy Selenium. *Phys. Rev. B* **2020**, *102*, 104202.
48. Tsuchiya, Y., Sound Velocity in the Liquid Tl-Se System. *J Non-Cryst Solids* **1993**, *156*, 700-703.
49. Tsuchiya, Y., Thermodynamics of the Structural Changes in the Liquid Ge-Te System around the Te-Rich Eutectic Composition. *J Non-Cryst Solids* **2002**, *312-14*, 212-216.
50. Akola, J.; Jones, R. O.; Kohara, S.; Usuki, T.; Bychkov, E., Density Variations in Liquid Tellurium: Roles of Rings, Chains, and Cavities. *Phys. Rev. B* **2010**, *81*, 094202.
51. Akola, J.; Jones, R. O., Structure and Dynamics in Amorphous Tellurium and Te-N Clusters: A Density Functional Study. *Phys. Rev. B* **2012**, *85*, 134103.
52. Sosso, G. C.; Colombo, J.; Behler, J.; Del Gado, E.; Bernasconi, M., Dynamical Heterogeneity in the Supercooled Liquid State of the Phase Change Material Gete. *J Phys Chem B* **2014**, *118*, 13621-13628.
53. Jara, D. A. C.; Michelon, M. F.; Antonelli, A.; de Koning, M., Theoretical Evidence for a First-Order Liquid-Liquid Phase Transition in Gallium. *Journal of Chemical Physics* **2009**, *130*, 221101.
54. Li, R. Z.; Sun, G.; Xu, L. M., Anomalous Properties and the Liquid-Liquid Phase Transition in Gallium. *Journal of Chemical Physics* **2016**, *145*, 054506.
55. Sun, P., et al., Structural Changes across Thermodynamic Maxima in Supercooled Liquid Tellurium: A Water-Like Scenario. *P Natl Acad Sci USA* **2022**, *119*, e2202044119.
56. Anikeenko, A. V.; Medvedev, N. N., Polytetrahedral Nature of the Dense Disordered Packings of Hard Spheres. *Phys. Rev. Lett.* **2007**, *98*, 235504.
57. Xia, C. J.; Li, J. D.; Cao, Y. X.; Kou, B. Q.; Xiao, X. H.; Fezzaa, K.; Xiao, T. Q.; Wang, Y. J., The Structural Origin of the Hard-Sphere Glass Transition in Granular Packing. *Nat. Commun.* **2015**, *6*, 8409.
58. Douglas, J. F.; Dudowicz, J.; Freed, K. F., Lattice Model of Equilibrium Polymerization. Vii. Understanding the Role of "Cooperativity" in Self-Assembly. *Journal of Chemical Physics* **2008**, *128*, 224901.
59. Shi, R.; Russo, J.; Tanaka, H., Common Microscopic Structural Origin for Water's Thermodynamic and Dynamic Anomalies. *Journal of Chemical Physics* **2018**, *149*, 224502.
60. Shi, R.; Russo, J.; Tanaka, H., Origin of the Emergent Fragile-to-Strong Transition in Supercooled Water. *P Natl Acad Sci USA* **2018**, *115*, 9444-9449.
61. Shi, R.; Tanaka, H., The Anomalies and Criticality of Liquid Water. *P Natl Acad Sci USA* **2020**, *117*, 26591-26599.
62. Yu, Z. H.; Shi, R.; Tanaka, H., A Unified Description of the Liquid Structure, Static and Dynamic Anomalies, and Criticality of Tip4p/2005 Water by a Hierarchical Two-State Model. *J Phys Chem B* **2023**, *127*, 3452-3462.
63. Douglas, J. F.; Dudowicz, J.; Freed, K. F., Does Equilibrium Polymerization Describe the Dynamic Heterogeneity of Glass-Forming Liquids? *Journal of Chemical Physics* **2006**, *125*, 144907.
64. Dudowicz, J.; Freed, K. F.; Douglas, J. F., Lattice Model of Equilibrium Polymerization. Iv. Influence of Activation, Chemical Initiation, Chain Scission and Fusion, and Chain Stiffness on Polymerization and Phase Separation. *Journal of Chemical Physics* **2003**, *119*, 12645-12666.
65. Tanaka, F.; Matsuyama, A., Tricriticality in Thermoreversible Gels. *Phys. Rev. Lett.* **1989**, *62*, 2759-2762.

66. Naserifar, S.; Goddard, W. A., Liquid Water Is a Dynamic Polydisperse Branched Polymer. *P Natl Acad Sci USA* **2019**, *116*, 1998-2003.
67. Neophytou, A.; Chakrabarti, D.; Sciortino, F., Topological Nature of the Liquid-Liquid Phase Transition in Tetrahedral Liquids. *Nature Physics, in press* **2022**, *18*, 1248-1253.
68. Betancourt, B. A. P.; Douglas, J. F.; Starr, F. W., String Model for the Dynamics of Glass-Forming Liquids. *Journal of Chemical Physics* **2014**, *140*, 204509.
69. Starr, F. W.; Douglas, J. F., Modifying Fragility and Collective Motion in Polymer Melts with Nanoparticles. *Phys. Rev. Lett.* **2011**, *106*, 115702.
70. Xu, W. S.; Douglas, J. F.; Xia, W. J.; Xu, X. L., Investigation of the Temperature Dependence of Activation Volume in Glass-Forming Polymer Melts under Variable Pressure Conditions. *Macromolecules* **2020**, *53*, 6828-6841.
71. Xu, W. S.; Douglas, J. F.; Xia, W. J.; Xu, X. L., Understanding Activation Volume in Glass-Forming Polymer Melts Via Generalized Entropy Theory. *Macromolecules* **2020**, *53*, 7239-7252.
72. Douglas, J. F.; Hubbard, J. B., Semiempirical Theory of Relaxation - Concentrated Polymer-Solution Dynamics. *Macromolecules* **1991**, *24*, 3163-3177.
73. Stukalin, E. B.; Douglas, J. F.; Freed, K. F., Multistep Relaxation in Equilibrium Polymer Solutions: A Minimal Model of Relaxation in "Complex" Fluids. *Journal of Chemical Physics* **2008**, *129*, 094901.
74. Douglas, J. F., Integral Equation Approach to Condensed Matter Relaxation. *Journal of Physics-Condensed Matter* **1999**, *11*, A329-A340.
75. Bernatz, K. M.; Echeverria, I.; Simon, S. L.; Plazek, D. J., Characterization of the Molecular Structure of Amorphous Selenium Using Recoverable Creep Compliance Measurements. *J Non-Cryst Solids* **2002**, *307*, 790-801.
76. Faivre, G.; Gardissat, J. L., Viscoelastic Properties and Molecular Structure of Amorphous Selenium. *Macromolecules* **1986**, *19*, 1988-1996.
77. Hagg, G., The Vitreous State. *Journal of Chemical Physics* **1935**, *3*, 42-49.
78. Zachariasen, W. H., The Atomic Arrangement in Glass. *J Am Chem Soc* **1932**, *54*, 3841-3851.
79. Douglas, J. F., Swelling and Growth of Polymers, Membranes, and Sponges. *Phys. Rev. E* **1996**, *54*, 2677-2689.
80. Crick, F. H. C.; Watson, J. D., Structure of Small Viruses. *Nature* **1956**, *177*, 473-475.
81. Caspar, D. L. D.; Klug, A., Physical Principles in the Construction of Regular Viruses. *Cold Spring Harbor Symposia on Quantitative Biology* **1962**, *27*, 1-24.
82. Klug, A., Architectural Design of Spherical Viruses. *Nature* **1983**, *303*, 378-379.
83. Caspar, D. L. D.; Fontano, E., Five-Fold Symmetry in Crystalline Quasicrystal Lattices. *P Natl Acad Sci USA* **1996**, *93*, 14271-14278.
84. Klug, A., Macromolecular Order in Biology. *Philos T R Soc A* **1994**, *348*, 167-178.
85. Zachariasen, W. H., The Vitreous State. *Journal of Chemical Physics* **1935**, *3*, 162-163.
86. Hagg, G., The Vitreous State. *Journal of Chemical Physics* **1935**, *3*, 363-364.
87. Tsuchiya, Y., Concentration Fluctuations Induced by Thermally-Driven Local Order Changes in the Molten Binary Alloy. *Scandinavian Journal of Metallurgy* **2001**, *30*, 345-351.
88. Kurita, R.; Murata, K.; Tanaka, H., Control of Fluidity and Miscibility of a Binary Liquid Mixture by the Liquid-Liquid Transition. *Nat Mater* **2008**, *7*, 647-652.
89. Dudowicz, J.; Douglas, J. F.; Freed, K. F., Self-Assembly in a Polymer Matrix and Its Impact on Phase Separation. *J Phys Chem B* **2009**, *113*, 3920-3931.
90. Tsuchiya, Y., The Thermodynamics of Structural Changes in the Liquid Sulphur-Tellurium System: Compressibility and Ehrenfest's Relations. *Journal of Physics-Condensed Matter* **1994**, *6*, 2451-2458.
91. Tsuchiya, Y.; Kakinuma, F.; Bergman, C., Structural Changes and Concentration Fluctuations in the Liquid Se-Te System. *J Non-Cryst Solids* **1996**, *205*, 143-146.
92. Coulet, M. V.; Bergman, C.; Bellissent, R.; Bichara, C., Local Order and Phase Separation in Sulphur-Tellurium Melts: A Neutron Scattering Study. *J Non-Cryst Solids* **1999**, *250*, 463-467.

93. Coulet, M.-V.; Bellissent, R.; Bichara, C., Closed-Loop Miscibility Gap in Sulfur-Tellurium Melts: Structural Evidence and Thermodynamic Modelling. *Journal of Physics-Condensed Matter* **2006**, *18*, 11471-11486.
94. Graf, E. H.; Lee, D. M.; Reppy, J. D., Phase Separation and the Superfluid Transition in Liquid He3-He4 Mixtures. *Phys. Rev. Lett.* **1967**, *19*, 417.
95. Leiderer, P.; Bosch, W., Universality of Tricritical He3-He4 Mixtures under Pressure. *Phys. Rev. Lett.* **1980**, *45*, 727-729.
96. Feynman, R. P., The Gamma-Transition in Liquid Helium. *Phys. Rev.* **1953**, *90*, 1116-1117.
97. Feynman, R. P., Atomic Theory of the 2-Fluid Model of Liquid Helium. *Phys. Rev.* **1954**, *94*, 262-277.
98. Zhang, H.; Zhong, C.; Douglas, J. F.; Wang, X. D.; Cao, Q. P.; Zhang, D. X.; Jiang, J. Z., Role of String-Like Collective Atomic Motion on Diffusion and Structural Relaxation in Glass Forming Cu-Zr Alloys. *Journal of Chemical Physics* **2015**, *142*, 164506.
99. Dudowicz, J.; Freed, K. F.; Douglas, J. F., Lattice Model of Living Polymerization. Ii. Interplay between Polymerization and Phase Stability. *Journal of Chemical Physics* **2000**, *112*, 1002-1010.
100. Zhou, Z.; Chu, B., Phase Behavior and Association Properties of Poly(Oxypropylene)-Poly(Oxyethylene)-Poly(Oxypropylene) Triblock Copolymer in Aqueous Solution. *Macromolecules* **1994**, *27*, 2025-2033.
101. Huff, A.; Patton, K.; Odhner, H.; Jacobs, D. T.; Clover, B. C.; Greer, S. C., Micellization and Phase Separation for Triblock Copolymer 17r4 in H2o and in D2o. *Langmuir* **2011**, *27*, 1707-1712.
102. Tan, H. M.; Moet, A.; Hiltner, A.; Baer, E., Thermoreversible Gelation of Atactic Polystyrene Solutions. *Macromolecules* **1983**, *16*, 28-34.
103. Kawanishi, K.; Takeda, Y.; Inoue, T., The Sol-Gel Transition and the Liquid-Liquid Phase Separation in Poly(Vinyl Chloride) Solutions. *Polym. J.* **1986**, *18*, 411-416.
104. Tanaka, H., Liquid–Liquid Transition and Polyamorphism. *Journal of Chemical Physics* **2020**, *153*, 130901.
105. Rah, K.; Freed, K. F.; Dudowicz, J.; Douglas, J. F., Lattice Model of Equilibrium Polymerization. V. Scattering Properties and the Width of the Critical Regime for Phase Separation. *Journal of Chemical Physics* **2006**, *124*, 144906.
106. Kumar, P.; Buldyrev, S. V.; Becker, S. R.; Poole, P. H.; Starr, F. W.; Stanley, H. E., Relation between the Widom Line and the Breakdown of the Stokes-Einstein Relation in Supercooled Water. *P Natl Acad Sci USA* **2007**, *104*, 9575-9579.
107. Simeoni, G. G.; Bryk, T.; Gorelli, F. A.; Krisch, M.; Ruocco, G.; Santoro, M.; Scopigno, T., The Widom Line as the Crossover between Liquid-Like and Gas-Like Behaviour in Supercritical Fluids. *Nature Physics, in press* **2010**, *6*, 503-507.
108. Xu, L. M.; Kumar, P.; Buldyrev, S. V.; Chen, S. H.; Poole, P. H.; Sciortino, F.; Stanley, H. E., Relation between the Widom Line and the Dynamic Crossover in Systems with a Liquid-Liquid Phase Transition. *P Natl Acad Sci USA* **2005**, *102*, 16558-16562.
109. Dudowicz, J.; Freed, K. F.; Douglas, J. F., Flory-Huggins Model of Equilibrium Polymerization and Phase Separation in the Stockmayer Fluid. *Phys. Rev. Lett.* **2004**, *92*, 045502.
110. Van Workum, K.; Douglas, J. F., Equilibrium Polymerization in the Stockmayer Fluid as a Model of Supermolecular Self-Organization. *Phys. Rev. E* **2005**, *71*, 031502.
111. Nishikawa, K.; Tanaka, I., Correlation Lengths and Density Fluctuations in Supercritical States of Carbon Dioxide. *Chemical Physics Letters* **1995**, *244*, 149-152.
112. Nishikawa, K.; Morita, T., Fluid Behavior at Supercritical States Studied by Small-Angle X-Ray Scattering. *J. Supercrit. Fluids* **1998**, *13*, 143-148.
113. Nishikawa, K.; Kusano, K.; Arai, A. A.; Morita, T., Density Fluctuation of a Van Der Waals Fluid in Supercritical State. *Journal of Chemical Physics* **2003**, *118*, 1341-1346.
114. Arai, A. A.; Morita, T.; Nishikawa, K., Analysis to Obtain Precise Density Fluctuation of Supercritical Fluids by Small-Angle X-Ray Scattering. *Chemical Physics* **2005**, *310*, 123-128.

115. Zhu, J. L.; Zhang, P. W.; Wang, H.; Delle Site, L., Is There a Third Order Phase Transition for Supercritical Fluids? *Journal of Chemical Physics* **2014**, *140*, 014502.
116. Angell, C. A., The Amorphous State Equivalent of Crystallization: New Glass Types by First Order Transition from Liquids, Crystals, and Biopolymers. *Solid State Sci.* **2000**, *2*, 791-805.
117. Angell, C. A.; Moynihan, C. T.; Hemmati, M., 'Strong' and 'Superstrong' Liquids, and an Approach to the Perfect Glass State Via Phase Transition. *J Non-Cryst Solids* **2000**, *274*, 319-331.
118. Torquato, S., Hyperuniform States of Matter. *Phys Rep* **2018**, *745*, 1-95.
119. Xu, W. S.; Douglas, J. F.; Freed, K. F., Influence of Cohesive Energy on the Thermodynamic Properties of a Model Glass-Forming Polymer Melt. *Macromolecules* **2016**, *49*, 8341-8354.
120. Angell, C. A., Two-State Thermodynamics and Transport Properties for Water from "Bond Lattice" Model. *Journal of Physical Chemistry* **1971**, *75*, 3698-&.
121. Moynihan, C. T.; Angell, C. A., Bond Lattice or Excitation Model Analysis of the Configurational Entropy of Molecular Liquids. *J Non-Cryst Solids* **2000**, *274*, 131-138.
122. Zhang, H.; Wang, X. Y.; Chremos, A.; Douglas, J. F., Superionic Uo2: A Model Anharmonic Crystalline Material. *Journal of Chemical Physics* **2019**, *150*, 174506.